\documentclass[aps,prd,a4paper,preprintnumbers,amsmath,amssymb,superscriptaddress,twocolumn,floatfix,showpacs,nofootinbib]{revtex4}
\usepackage{amssymb}
\usepackage{float}
\usepackage{graphicx}
\usepackage{epstopdf}

\begin{document}
\title{Structure Formation by Fifth Force I:  N-Body vs. Linear Simulations}
\author{Baojiu~Li}
\email[Email address: ]{b.li@damtp.cam.ac.uk} \affiliation{DAMTP,
Centre for Mathematical Sciences, University of Cambridge,
Cambridge CB3 0WA, UK}
\affiliation{Kavli Institute for Cosmology
Cambridge, Madingley Road, Cambridge CB3 0HA, UK}
\author{Hongsheng~Zhao}
\email[Email address: ]{hz4@st-andrews.ac.uk} \affiliation{SUPA,
University of St.~Andrews, North Haugh, Fife, KY16 9SS, UK}
\affiliation{Leiden Observatory, Leiden University, Niels Bohrweg
2, Leiden, The Netherlands }
\date{\today}

\begin{abstract}
We lay out the frameworks to numerically study the structure
formation in both linear and nonlinear regimes in general
dark-matter-coupled scalar field models, and give an explicit
example where the scalar field serves as a dynamical dark energy.
Adopting parameters of the scalar field which yield a realistic
CMB spectrum, we generate the initial conditions for our N-body
simulations, which follow the spatial distributions of the dark
matter and the scalar field by solving their equations of motion
using the multilevel adaptive grid technique. We show that the
spatial configuration of the scalar field tracks well the voids
and clusters of dark matter. Indeed, the propagation of scalar
degree of freedom effectively acts a fifth force on dark matter
particles, whose range and magnitude are determined by the two
model parameters $(\mu, \gamma)$, local dark matter density as
well as the background value for the scalar field. The model
behaves like the $\Lambda$CDM paradigm on scales relevant to the
CMB spectrum, which are well beyond the probe of the local fifth
force and thus not significantly affected by the matter-scalar
coupling. On scales comparable or shorter than the range of the
local fifth force, the fifth force is perfectly parallel to
gravity and their strengths have a fixed ratio $2\gamma^{2}$
determined by the matter-scalar coupling, provided that the
chameleon effect is weak; if on the other hand there is a strong
chameleon effect (\emph{i.e.}, the scalar field almost resides at
its effective potential minimum everywhere in the space), the
fifth force indeed has suppressed effects in high density regions
and shows no obvious correlation with gravity, which means that
the dark-matter-scalar-field coupling is not simply equivalent to
a rescaling of the gravitational constant or the mass of the dark
matter particles. We show these spatial distributions and (lack
of) correlations at typical redshifts ($z = 0, 1, 5.5$) in our
multi-grid million-particle simulations. The viable parameters for
the scalar field can be inferred on intermediate or small scales
at late times from, \emph{e.g.}, weak lensing and phase space
properties, while the predicted Hubble expansion and linearly
simulated CMB spectrum are virtually indistinguishable from the
standard $\Lambda$CDM predictions.
\end{abstract}


\maketitle

\section{Introduction}

\label{sect:intro}

The origin and nature of dark energy \cite{Copeland2006} is one of
the most difficult challenges facing physicists and cosmologists
now. Among all the proposed models to tackle this problem, a
scalar field is perhaps the most popular one up to now. The scalar
field, denoted by $\varphi$, might only interact with other matter
species through gravity, or have a coupling to normal matter and
therefore producing a fifth force on matter particles. The latter
idea has seen a lot of interests in recent years, in the light
that such a coupling could alleviate the coincidence problem of
dark energy and that it is commonly predicted by low energy
effective theories from a fundamental theory.

Nevertheless, if there is a coupling between the scalar field and
baryonic particles, then stringent experimental constraints might
be placed on the fifth force on the latter provided that the
scalar field mass is very light (which is needed for the dark
energy). Such constraints severely limit the viable parameter
space of the model. Different ways out of the problem have been
proposed, of which the simplest one is to have the scalar field
coupling to dark matter only but not to standard model particles,
therefore evading those constraints entirely. This is certainly
possible, especially because both dark matter and dark energy are
unknown to us and they may well have a common origin. Another
interesting possibility is to have the chameleon mechanism
\cite{Khoury2004a, Khoury2004b, Mota2006, Mota2007}, by virtue of
which the scalar field acquires a large mass in high density
regions and thus the fifth force becomes undetectablly
short-ranged, and so also evades the constraints.

Study of the cosmological effect of a chameleon scalar field shows
that the fifth force is so short-ranged that it has negligible
effect in the large scale structure formation \cite{Brax2004} for
certain choices of the scalar field potential. But it is possible
that the scalar field has a large enough mass in the solar system
to pass any constraints, and at the same time has a low enough
mass (thus long range forces) on cosmological scales, producing
interesting phenomenon in the structure formation. This is the
case of some $f(R)$ gravity models \cite{Carroll2004, Nojiri2003},
which survives solar system tests thanks again to the chameleon
effect \cite{Navarro2007, Li2007, Hu2007, Brax2008}. Note that the
$f(R)$ gravity model is mathematically equivalent to a scalar
field model with matter coupling.

No matter whether the scalar field couples with dark matter only
or with all matter species, it is of general interests to study
its effects in cosmology, especially in the large scale structure
formation. Indeed, at the linear perturbation level there have
been a lot of studies about the coupled scalar field and $f(R)$
gravity models which enable us to have a much clearer picture
about their behaviors now. But linear perturbation studies do not
conclude the whole story, because it is well known that the matter
distribution at late times becomes nonlinear, making the behavior
of the scalar field more complex and the linear analysis
insufficient to produce accurate results to confront with
observations. For the latter purpose the best way is to perform
full N-body simulations \cite{Bertschinger1998} to evolve the
individual particles step by step.

N-body simulations for scalar field and relevant models have been
performed before \cite{Linder2003, Mainini2003, Springel2007,
Kesden2006, Farrar2007, Keselman2009, Maccio2004, Baldi2008}. For
example, in \cite{Maccio2004} the simulation is about a specific
coupled scalar field model. This study however does not obtain a
full solution to the spatial configuration of the scalar field,
but instead simplifies the simulation by assuming that the scalar
field's effect is to change the value of the gravitational
constant, and presenting an justifying argument for such an
approximation. As we will see below, this approximation is only
good in certain parameter spaces and for certain choices of the
scalar field potential, and therefore we believe fuller
simulations than the one performed in \cite{Maccio2004} is needed
to study the scalar field behavior most rigorously. Note also the
structure formation with a coupled chameleon scalar field has also
been studied using non B-body techniques previously
\cite{Brax2006}.

Recently there have also appeared full simulations of the $f(R)$
gravity model \cite{Oyaizu2008, Oyaizu2008b}, which do solve the
scalar degree of freedom explicitly. However, embedded in the
$f(R)$ framework there are some limitations in the generality of
these works. As a first thing, $f(R)$ gravity model (no matter
what the form $f$ is) only corresponds to the couple scalar field
models for a specific value of coupling strength
\cite{Amendola2007}. Second, in $f(R)$ models the correction to
standard general relativity is through the modification to the
Poission equation and thus to the gravitational potential as a
whole \cite{Oyaizu2008}, while in the coupled scalar field models
we could clearly separate the scalar fifth force from gravity and
analyze the former directly. Also, in $f(R)$ models the coupling
between matter and the scalar field is universal (the same to dark
matter and baryons), while in the couple scalar field models it is
straightforward to switch on/off the coupling to baryons and study
the effects on baryonic and dark matter clusterings respectively.
And finally, the general framework of N-body simulations in couple
scalar field models could also handle the situation where the
chameleon effect is absent and scalar field only couples to dark
matter (which is of no interests for $f(R)$ people).

In this article we present the general formulae and algorithm of
full N-body simulations in coupled scalar field models and
consider as an explicit example the results for a chameleon scalar
field. Unlike in \cite{Maccio2004}, here we shall calculate the
spatial distribution of the scalar field directly without making
simplifications such as a rescaled gravitational constant. Neither
shall we use the concept of varying-mass dark matter particles as
in \cite{Maccio2004}, but instead we treat the system as
constant-mass dark matter particles under the action of a fifth
force.

The article is organized as follows: in \S~\ref{sect:model} we
review the general equations of motion for the coupled scalar
field model and introduce our specific choices of the coupling
function and scalar field potential. Next we analyze in
\S~\ref{sect:lin} the general linear perturbation equations in the
$3+1$ formalism \cite{Ellis1989, Challinor1999} and integrate them
into the numerical Boltzmann code CAMB \cite{Lewis2000} to study
the effects on the linear structure formation. Then in
\S~\ref{sect:nonlin} we turn to our main focus, introducing the
formulae and algorithm of the N-body simulation. We also study the
chosen model explicitly, display the preliminary results and
discuss on them. Finally we conclude in \S~\ref{sect:disc}.

\section{The Coupled Scalar Field Model}

\label{sect:model}

In this section we briefly introduce the model considered here and
present the equations that will be analyzed in the following
sections. Let us start by looking at the general field equations
for a scalar field coupled to dark matter.

The Lagrangian for our coupled scalar field model is
\begin{eqnarray}\label{eq:Lagrangian}
\mathcal{L} &=&
\frac{1}{2}\left[\frac{R}{\kappa}-\nabla^{a}\varphi\nabla_{a}\varphi\right]
+ V(\varphi) - C(\varphi)\mathcal{L}_{\mathrm{CDM}} +
\mathcal{L}_{\mathrm{S}}\ \
\end{eqnarray}
where $R$ is the Ricci scalar, $\kappa=8\pi G$ with $G$ Newton's
constant, $\varphi$ is the scalar field, $V(\varphi)$ is its
potential energy and $C(\varphi)$ its coupling to dark matter,
which is assumed to be cold and described by the Lagrangian
$\mathcal{L}_{\mathrm{CDM}}$. $\mathcal{L}_{\mathrm{S}}$ includes
all the terms for photons, neutrinos and baryons, and these will
be considered only when we calculate the large scale structure
formation in the next section.

The dark matter Lagrangian, for a point-like particle with (bare)
mass $m_{0}$, is
\begin{eqnarray}\label{eq:DMLagrangian}
\mathcal{L}_{\mathrm{CDM}}(\mathbf{y}) &=&
-\frac{m_{0}}{\sqrt{-g}}\delta(\mathbf{y}-\mathbf{x}_{0})\sqrt{g_{ab}\dot{x}^{a}_{0}\dot{x}^{b}_{0}}
\end{eqnarray}
where $\mathbf{y}$ is the coordinate and $\mathbf{x}_{0}$ is the
coordinate of the centre of the particle. From this equation it
can be easily derived that
\begin{eqnarray}\label{eq:DMEMT_particle}
T^{ab}_{\mathrm{CDM}} &=&
\frac{m_{0}}{\sqrt{-g}}\delta(\mathbf{y}-\mathbf{x}_{0})
\dot{x}^{a}_{0}\dot{x}^{b}_{0}.
\end{eqnarray}
Also, because $g_{ab}\dot{x}^{a}_{0}\dot{x}^{b}_{0} =
g_{ab}u^{a}u^{b}=1$ where $u^{a}$ is the four velocity of the dark
matter particle, so the Lagrangian could be rewritten as
\begin{eqnarray}\label{eq:DMLagrangian2}
\mathcal{L}_{\mathrm{CDM}}(\mathbf{y}) &=&
-\frac{m_{0}}{\sqrt{-g}}\delta(\mathbf{y}-\mathbf{x}_{0}),
\end{eqnarray}
which will be used below.

\begin{figure*}[tbp]
\centering \includegraphics[scale=1.2] {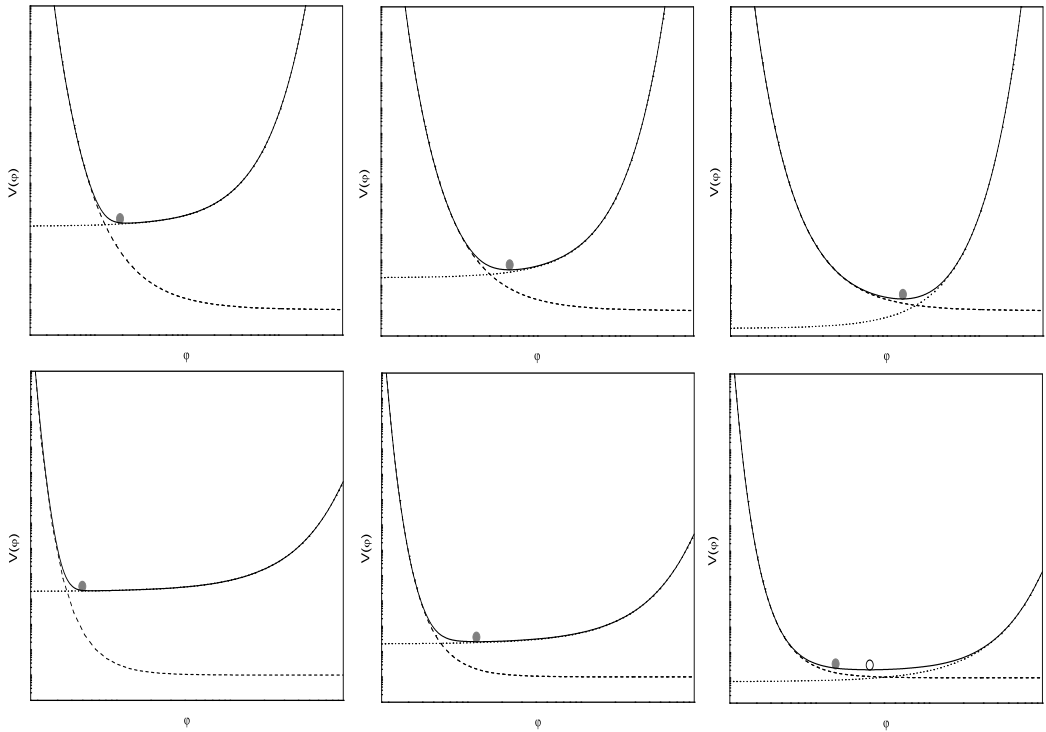}
\caption{Upper panel: the potential (dashed curves), coupling
function (dotted curves) and effective potential (solid curves) of
the strong coupling scalar field model at various cosmic epochs.
Lower panel: the same as the upper panel but for a coupling not
strong enough so that the scalar field will not reside at the
minimum of $V_{eff}$. Solid circles denote the states of the
scalar field and in case that this does not coincide with the
minimum of effective potential, the later is indicated by an open
circle. See text for explanations.} \label{fig:Figure1}
\end{figure*}

Eq.~(\ref{eq:DMEMT_particle}) is the energy momentum tensor for a
single dark matter particle. For a fluid with many particles the
energy momentum tensor will be
\begin{eqnarray}\label{eq:DMEMT_fluid}
T^{ab}_{\mathrm{CDM}} &=& \frac{1}{V}\int_{V}d^{4}y\sqrt{-g}
\frac{m_{0}}{\sqrt{-g}}\delta(y-x_{0})
\dot{x}^{a}_{0}\dot{x}^{b}_{0}\nonumber\\
&=& \rho_{\mathrm{CDM}}u^{a}u^{b},
\end{eqnarray}
in which $V$ is a volume microscopically large and macroscopically
small, and we have extend the 3-dimensional $\delta$ function to a
4-dimensional one by adding a time component.

Meanwhile, using
\begin{eqnarray}
T^{ab} &=&
-\frac{2}{\sqrt{-g}}\frac{\delta\left(\sqrt{-g}\mathcal{L}\right)}{\delta
g_{ab}}
\end{eqnarray}
it is straightforward to show that the energy momentum tensor for
the scalar field is given by
\begin{eqnarray}\label{eq:phiEMT}
T^{\varphi ab} &=& \nabla^{a}\varphi\nabla^{b}\varphi -
g^{ab}\left[\frac{1}{2}\nabla_{c}\varphi\nabla^{c}\varphi -
V(\varphi)\right].
\end{eqnarray}
So the total energy momentum tensor is
\begin{eqnarray}\label{eq:EMT_tot}
T_{ab} &=& \nabla_{a}\varphi\nabla_{b}\varphi -
g_{ab}\left[\frac{1}{2}\nabla_{c}\varphi\nabla^{c}\varphi -
V(\varphi)\right]\nonumber\\
&& + C(\varphi)T^{\mathrm{CDM}}_{ab} + T^{\mathrm{S}}_{ab}
\end{eqnarray}
where $T^{\mathrm{CDM}}_{ab}=\rho_{\mathrm{CDM}}u_{a}u_{b}$,
$T^{\mathrm{S}}_{ab}$ is the energy momentum tensor for standard
model particles, and the Einstein equation is
\begin{eqnarray}\label{eq:EinsteinEq}
G_{ab} &=& \kappa T_{ab}
\end{eqnarray}
where $G_{ab}$ is the Einstein tensor. Note that due to the
coupling between the scalar field $\varphi$ and the dark matter,
the energy momentum tensors for either will not be conserved, and
we have
\begin{eqnarray}\label{eq:DM_energy_conservation}
\nabla_{b}T^{\mathrm{CDM}_{}ab} &=&
-\frac{C_{\varphi}(\varphi)}{C(\varphi)}\left(g^{ab}\mathcal{L}_{\mathrm{CDM}}
+ T^{\mathrm{CDM}ab}\right)\nabla_{b}\varphi\ \
\end{eqnarray}
where throughout this paper we shall use a $_{\varphi}$ to denote
the derivative with respect to $\varphi$.

Finally, the scalar field equation of motion (EOM) from the given
Lagrangian is
\begin{eqnarray}
\square\varphi + \frac{\partial V(\varphi)}{\partial\varphi} &=&
\frac{\partial
C(\varphi)}{\partial\varphi}\mathcal{L}_{\mathrm{CDM}}\nonumber
\end{eqnarray}
where $\square = \nabla^{a}\nabla_{a}$. Using
Eq.~(\ref{eq:DMLagrangian2}) it can be rewritten as
\begin{eqnarray}\label{eq:phiEOM}
\square\varphi + \frac{\partial V(\varphi)}{\partial\varphi} +
\rho_{\mathrm{CDM}}\frac{\partial C(\varphi)}{\partial\varphi} &=&
0.
\end{eqnarray}

Eqs.~(\ref{eq:EMT_tot}, \ref{eq:EinsteinEq},
\ref{eq:DM_energy_conservation}, \ref{eq:phiEOM}) summarize all
the physics that will be used in our analysis.

We will consider a special form for the scalar field potential,
\begin{eqnarray}\label{eq:potential}
V(\varphi) &=& \frac{V_{0}}
{\left[1-\exp\left(-\sqrt{\kappa}\varphi\right)\right]^{\mu}},
\end{eqnarray}
where we have fixed the coefficient in front of $\varphi$ to be
$1$ without loss of generality, since we can always rescale
$\varphi$ to achieve this; $\mu$ is a dimensionless constant and
$V_{0}$ is a constant with mass dimension $4$. As will be
discussed below, we need $\mu\ll1$ to evade local observational
constraints. Meanwhile, the coupling between the scalar field and
dark matter particle is chosen as
\begin{eqnarray}\label{eq:coupling_function}
C(\varphi) &=& \exp(\gamma\sqrt{\kappa}\varphi),
\end{eqnarray}
where $\gamma>0$ is another dimensionless constant.

As will be explained below, the two dimensionless parameters $\mu$
and $\gamma$ have clear physical meanings: \emph{roughly
speaking}, $\mu$ controls the time when the effect of the scalar
field becomes important in cosmology while $\gamma$ determines how
important it would ultimately be. Indeed, the potential given in
Eq.~(\ref{eq:potential}) is motivated by the $f(R)$ cosmology
\cite{Li2007}, in which the extra degree of freedom behaves as a
coupled scalar field in the Einstein frame. As we could see from
Eq.~(\ref{eq:potential}), the potential $V\rightarrow\infty$ when
$\varphi\rightarrow0$ while $V\rightarrow V_{0}$ when
$\varphi\rightarrow\infty$. In the latter case, however,
$C\rightarrow\infty$, so that the effective total potential
\begin{eqnarray}
V_{eff}(\varphi) &=& V(\varphi) + \rho_{\mathrm{CDM}}C(\varphi)
\end{eqnarray}
has a global minimum at some finite $\varphi$. If the total
potential $V_{eff}(\varphi)$ is steep enough around this minimum,
then the scalar field becomes very heavy and thus follows its
minimum dynamically, as is in the case of the chameleon cosmology.
If $V_{eff}$ is not steep enough at the minimum, however, the
scalar field will have a more complicated evolution. These two
different cases can be obtained by choosing appropriate values of
$\gamma, \mu$: if $\gamma$ is very large or $\mu$ is small then we
run into the former situation and if $\gamma$ is small and $\mu$
is large we have the latter.

In Fig.~\ref{fig:Figure1} we present a schematic plot of the two
situations: the significant difference is that at late times when
$\rho_{\mathrm{CDM}}$ becomes small, the effective potential
becomes flat around its minimum if $\gamma$ is not very large and
$\mu$ not very small, and as a result the true value of $\varphi$
will lag behind that corresponding to the minimum of $V_{eff}$. Of
course if $V_{0}$ is chosen appropriately the scalar field can act
as a dynamical dark energy in this slow-roll regime.

The complexity of the two cases also makes them phenomenologically
rich, and it is of our interests to study how such a setup will
affect the cosmology in background, linear perturbation and in
particular nonlinear structure formation regimes. In the next two
sections we shall consider these questions.

\begin{figure*}[tbp]
\centering
\includegraphics[scale=1.3] {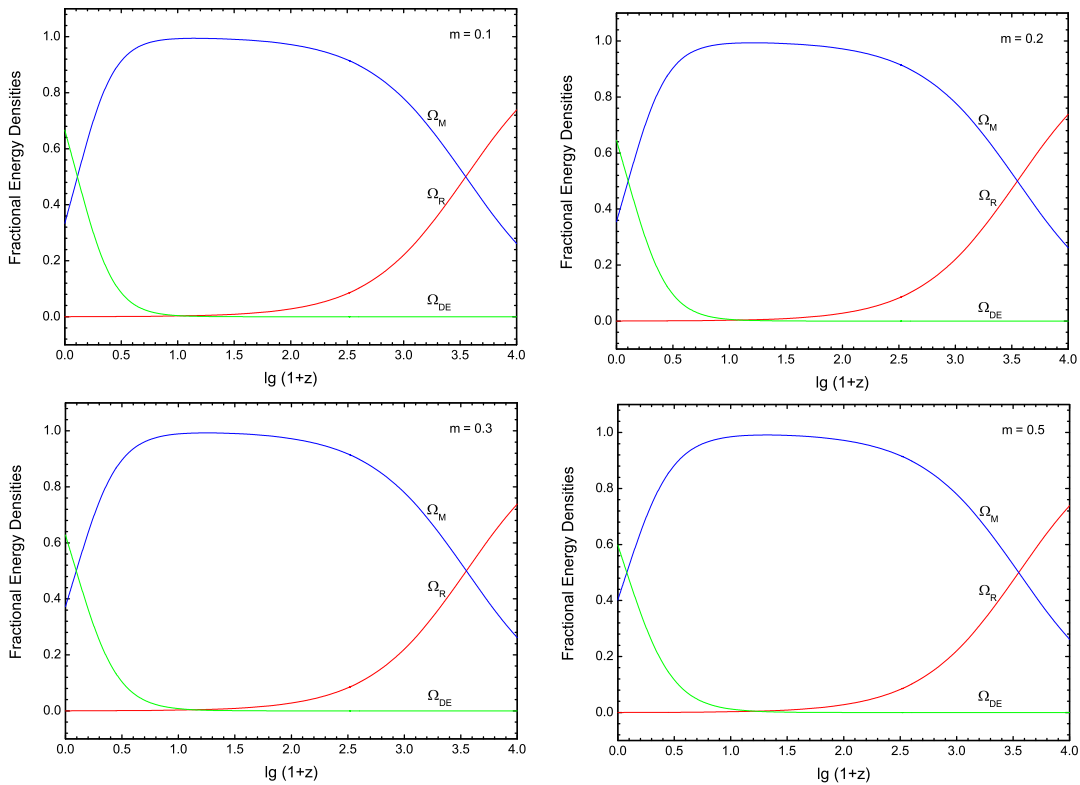}
\caption{(Color Online) The evolution of the fractional energy
densities in radiation (red lines), baryons + cold dark matter
(blue lines) and scalar field dark energy component (green lines)
with respect to $\lg(1+z)$ where $z$ is the redshift. The model
parameters for this case are chosen by $\gamma = 1$ and $\mu =
0.1, 0.2, 0.3, 0.5$ respectively. The physical parameters are
$\Omega_{\mathrm{R}}=8.475\times10^{-5}$ for radiation,
$\Omega_{\mathrm{B}}=0.05$ for baryons,
$\Omega_{\mathrm{\mathrm{CDM}}}=0.25$ for CDM and $H_{0} =
70~\mathrm{km/s/Mpc}$ for the present Hubble expansion rate. To
ensure these parameters the value of $V_{0}$ must be fine-tuned,
and here we have used a trial-and-error method to adjust the value
of $\lambda=\kappa V_{0}/3H_{0}^{2}$ as $0.53285, 0.44868,
0.38968, 0.30848$ respectively for $\mu=0.1, 0.2, 0.3, 0.5$. Note
that given $H_{0}$ the propagations of the fractional energy
densities fix the background evolution.} \label{fig:Figure2}
\end{figure*}

\begin{figure}[tbp]
\centering
\includegraphics[scale=0.85] {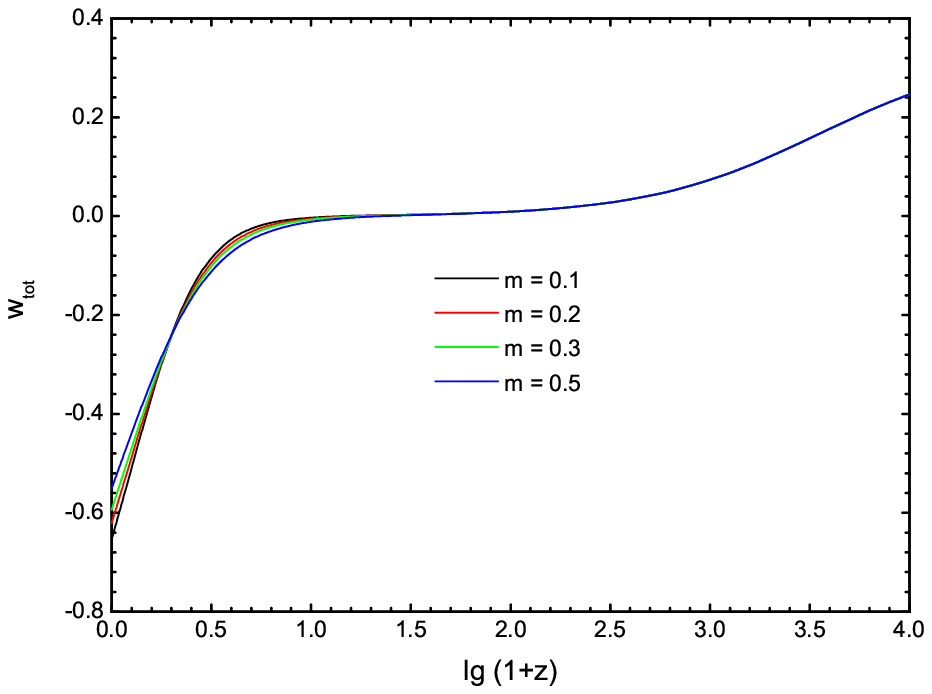}
\caption{(Color Online) The total equation of state of the model
with different values of $\mu$ as a function of $\lg(1+z)$. The
parameters are the same as in Fig.~\ref{fig:Figure2} and the
values of $\mu$ are given beneath the curves.} \label{fig:Figure3}
\end{figure}

\section{Linear Structure Formation}

\label{sect:lin}

Our first task is to linearize the model. Since this involves many
equations, we refer the interested readers to the Appendices
\ref{appen:perteqn} and \ref{appen:k-eqn} for the complete
first-order perturbation equations and their representations in
$k$-space. We shall proceed then to study their effects on the
large scale structure at linear perturbation level in the Universe
by putting these linearize equations into the numerical Boltzmann
code.

\subsection{The Background Evolution}

\label{subsect:bgd}

In what follows we will consider the limit of small $\mu$,
$\mu\lesssim\mathcal{O}(0.1)$. It is then useful to define the
effective mass of the scalar field by the Taylor expansion
\begin{eqnarray}
V(\varphi) &\approx& V(\varphi_{0}) + {m_{eff}^2 \over 2} (\varphi-\varphi_0)^2
\end{eqnarray}
where $\varphi_{0}$ is at the minimum of $V_{eff}$, given by
\begin{eqnarray}\label{eq:phi_min}
{dV_{eff}(\varphi_{0})\ \over\ d\varphi}\ =\ 0 &\rightarrow&
\sqrt{\kappa}\varphi_{0}\ \approx\ \frac{\mu}{\gamma}
\frac{V_{0}}{\rho_{\mathrm{CDM}}},
\end{eqnarray}
in which we have used the facts that $1+\mu\approx1$ as well as
$\sqrt{\kappa}\varphi\ll1$ so that
$\exp(-\sqrt{\kappa}\varphi)\approx\exp(\gamma\sqrt{\kappa}\varphi)\approx1$
and $1-\exp(-\sqrt{\kappa}\varphi)\approx\sqrt{\kappa}\varphi$.
Thus the effective mass
\begin{eqnarray}\label{eq:V_phiphi}
&& m^{2}_{eff}\nonumber\\ &\equiv&
\frac{\partial^{2}V_{eff}(\varphi_{0})}{\partial\varphi^{2}_0},  \quad y \equiv \exp(-\sqrt{\kappa}\varphi_0)\\
&=& \mu\frac{\kappa V_{0} y }  {\left[1- y \right]^{1+\mu}} \left[ 1 + { (1+\mu )y\over 1-y} \right]
+\gamma^{2}\kappa\rho_{\mathrm CDM} y^{-\gamma} \nonumber\\
&\approx& \frac{ (\gamma \kappa \rho_{\mathrm CDM})^2 }{\kappa V_{0}}
\left[{1+\mu \over \mu} + \ln { \mu V_0 \over \gamma \rho_{\mathrm CDM} }
+ {(1+3\gamma) V_0 \over \gamma \rho_{\mathrm CDM}}\right], \nonumber\\
&\approx& \frac{ (\gamma \kappa \rho_{\mathrm CDM})^2 }{\mu \kappa V_{0}}
\end{eqnarray}

To see that the scalar field is heavy, \emph{i.e.},
$m^{2}_{eff}\gg H^2 \sim H_0^{2}a^{-3}$, note that
$\kappa\rho_{\mathrm{CDM}}\sim\mathcal{O}\left(H_0^{2}a^{-3}\right)
\gg \mu \kappa V_0$, true for $\mu \ll 1$ [or  $\mu \sim \mathcal
O(1)$ but in the early universe, remember that we always assume
$\gamma\sim\mathcal{O}(1)$]. This means that for $\mu\ll1$ (or in
the early times) the scalar field has a very heavy mass and tends
to settle at $\varphi=\varphi_{0}$ (quickly oscillating there).
Then as a good approximation we have
\begin{eqnarray}
V(\varphi)\ \approx\ V(\varphi_{0})\ \approx\
V_{0}\left(\frac{\gamma\rho_{\mathrm{CDM}}}{V_{0}}\right)^{\mu}\left(\frac{1}{\mu}\right)^{\mu}\
\approx\ V_{0}\
\end{eqnarray}
in which we have used the facts that
$\lim_{\mu\rightarrow0^{+}}\mu^{\mu}=1$ in this regime. At the
same time, because
$\sqrt{\kappa}\dot{\varphi}\approx\sqrt{\kappa}\dot{\varphi_{0}}\approx
-\frac{\mu}{\gamma}\frac{V_{0}}{\rho_{\mathrm{CDM}}}
\frac{\dot{\rho}_{\mathrm{CDM}}}{\rho_{\mathrm{CDM}}}=3\sqrt{\kappa}\varphi_{0}H
\ll H$ with the use of Eq.~(\ref{eq:phi_min}), so the kinetic
energy of the scalar field is negligible. This shows that for
small $\mu$ we do expect the scalar field to behave like a
cosmological constant in background cosmology.

This analysis has been confirmed by numerical calculations. In
Fig.~\ref{fig:Figure2} we show the fractional energy densities of
the radiation, dust (baryon plus CDM) and dark energy (the scalar
field) components for $\gamma=1$ and several values of $\mu$. It
can be seen there the evolutions of these fractional energy
densities are not sensitive to $\mu$. To see this more clearly we
have also plotted in Fig.~\ref{fig:Figure3} the total equation of
state of all matter species. As $\mu\rightarrow0$ the curves
quickly become indistinguishable from each other (and
indistinguishable from the $\Lambda$CDM prediction).

\subsection{The Linear Perturbation Evolution}

\label{subsect:linper}

To study the perturbation evolution and the effect of the scalar
field on large scale structure formation, we just need to
implement the perturbation equations listed in
Appendix~\ref{appen:k-eqn} into a Boltzmann code. As mentioned
earlier, here we work in the $A=0$ frame, in which the CDM
peculiar velocity $v_{\mathrm{CDM}}$ must be dynamically evolved
according to Eq.~(\ref{eq:pb_v_CDM}) with $A=0$. Also note that
the tight coupling approximation is not affected as compared with
GR+$\Lambda$CDM.

The initial condition for $v_{\mathrm{CDM}}$ could be set to zero.
In this way the observer is comoving with the dark matter
particles initially (when the $\xi$ term in
Eq.~(\ref{eq:bg_rho_CDM}) is extremely small) and only deviate
from the dark matter particle world lines when this $\xi$ term
becomes significant. As for the initial conditions of $\xi$ and
$\xi'$, they are also very small at early times, and as we shall
see below, there is good reason to set them to zero at those
times. Thus we have adopted
\begin{eqnarray}\label{eq:initcond}
\xi_{\mathrm{initial}} &=& 0,\\
\xi'_{\mathrm{initial}} &=& 0,\\
v_{\mathrm{CDM, initial}} &=& 0,
\end{eqnarray}
as the initial conditions of the new variables which need to be
evolved in the code.

\begin{figure}[tbp]
\centering
\includegraphics[scale=0.85] {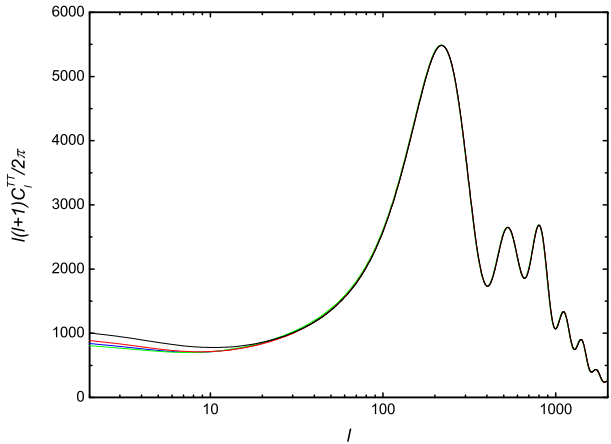}
\caption{(Color Online) The CMB power spectra for the couple
scalar field model with $\gamma=0.5$. The red, blue, green and
black curves correspond to $\mu = 0.1, 0.2, 0.3$ with chameleon
perturbations and $\mu = 0.1$ \emph{without} chameleon
perturbations, respectively. Curves for $\mu\lesssim10^{-5}$ are
indistinguishable from the $\Lambda$CDM result.}
\label{fig:Figure4}
\end{figure}

\begin{figure}[tbp]
\centering
\includegraphics[scale=0.85] {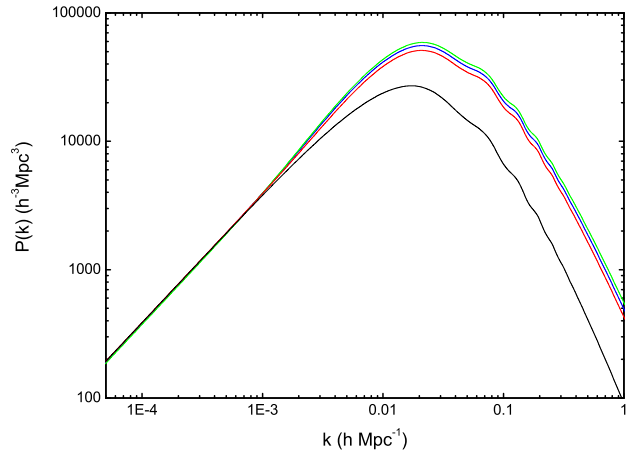}
\caption{(Color Online) The matter power spectra for the couple
scalar field model. The parameters are the same as in
Fig.~\ref{fig:Figure4}. The curves with $\mu\lesssim10^{-5}$ are
very close to the $\Lambda$CDM result.} \label{fig:Figure5}
\end{figure}

In Fig.~\ref{fig:Figure4} we show a collect of the CMB power
spectra for the model with $\gamma = 0.5$. In
Fig.~\ref{fig:Figure5} the matter power spectra of the
corresponding parameter choices are given. We can see that the
effect of the scalar field on the CMB spectrum is most significant
for low-$\ell$ (large scales), and even there the effect is only
to slightly reduce the power. This behavior is quite similar to
the one found in \cite{Li2007} and is due to the less decay of the
gravitational potential on large scales at late times, which
decreases the integrated Sachs-Wolfe (ISW) effect. Such small
deviations of the CMB spectrum from the $\Lambda$CDM result are
obviously not of much help in placing constraints on the model
parameters $\mu$ and $\gamma$, especially because of the cosmic
variance on large scales. Thus to distinguish this model from
others we necessarily need to use other observables such as the
linear (on large scales) and nonlinear (on smaller scales) matter
power spectra.

To explain the matter power spectrum, let us consider the
evolution of cold dark matter density contrast
$\Delta_{\mathrm{CDM}}$ (A similar analysis is first given in
\cite{Li2007} for $f(R)$ model and later more generally in
\cite{Tsujikawa2007}). For simplicity we shall assume a Universe
filled with only radiation and CDM particles. Taking the
(conformal) time derivative of the equation
\begin{eqnarray}\label{eq:DeltaEvolution1}
\Delta'_{\mathrm{CDM}} + k\mathcal{Z} + kv_{\mathrm{CDM}} &=& 0,
\end{eqnarray}
we have
\begin{eqnarray}\label{eq:DeltaEvolution2}
\Delta''_{\mathrm{CDM}} + k\mathcal{Z}' + kv'_{\mathrm{CDM}} &=&
0.
\end{eqnarray}
Taking the spatial derivative of Eq.~(\ref{eq:Raychaudhrui}),
collecting the coefficients of harmonic expansions and using the
frame choice $A = 0$ we obtain
\begin{eqnarray}\label{eq:ZEvolution}
k\mathcal{Z}' + k\frac{a'}{a}\mathcal{Z} +
\frac{\kappa}{2}(\mathcal{X}+3\mathcal{X}^{p})a^{2} &=& 0.
\end{eqnarray}
Substituting Eq.~(\ref{eq:ZEvolution}) into
Eq.~(\ref{eq:DeltaEvolution2}) and using Eqs.~(\ref{eq:pb_v_CDM},
\ref{eq:DeltaEvolution1}) we arrive at
\begin{eqnarray}\label{eq:DeltaEvolution3}
\Delta''_{\mathrm{CDM}} + \frac{a'}{a}\Delta'_{\mathrm{CDM}} -
\frac{\kappa}{2}(\mathcal{X}+3\mathcal{X}^{p})a^{2}\nonumber\\
+ k\frac{C_{\varphi}}{C}(k\xi - \varphi'v_{\mathrm{CDM}}) &=& 0,
\end{eqnarray}
where the last term is equal to $k\left(k A +
v'_{\mathrm{CDM}}+\frac{a'}{a}v_{\mathrm{CDM}}\right)$ with $\xi$
being substituted using the propagation equation of $v_{\mathrm{
CDM}}$ (remember that $A=0$ by choice of the frame).

Now according to Eq.~(\ref{eq:sfpert_EOM}) the evolution of $\xi$
is governed by
\begin{eqnarray}\label{eq:sfpert_EOM2}
\xi'' + 2\frac{a'}{a}\xi' + (k^{2} + a^{2}V_{\varphi\varphi} +
a^{2}\rho_{\mathrm{CDM}}C_{\varphi\varphi})\xi\nonumber\\ +
k\varphi'\mathcal{Z} +
a^{2}C_{\varphi}\rho_{\mathrm{CDM}}\Delta_{\mathrm{CDM}} &=& 0.\ \
\ \
\end{eqnarray}
For small scales (and late times) $k \gg |\frac{a'}{a}|$ and the
term $k^{2}\xi$ dominates over other terms proportional to $\xi$
in the above equations dominates so that the equation is
approximately
\begin{eqnarray}\label{eq:sf_appox1}
k^{2}\xi + a^{2}C_{\varphi}\rho_{\mathrm{CDM}}\Delta_{\mathrm{CDM}} &\doteq&
-k\varphi'\mathcal{Z} \\
&=& \varphi'\Delta'_{\mathrm{CDM}}
+ k\varphi'v_{\mathrm{CDM}}.
\label{eq:sf_pert}
\end{eqnarray}
where we choose to eliminate $\mathcal{Z}$ by using
Eq.~(\ref{eq:DeltaEvolution1}). Substituting
Eq.~(\ref{eq:sf_pert}) into Eq.~(\ref{eq:DeltaEvolution3}) to
eliminate $\xi$ and $v_{\mathrm{CDM}}$,  we get
\begin{eqnarray}
\Delta''_{\mathrm{CDM}} + \left(\frac{a'}{a} +
\frac{C_{\varphi}}{C}\varphi'\right)\Delta'_{\mathrm{CDM}}\nonumber\\
- \frac{\kappa}{2}(\mathcal{X}+3\mathcal{X}^{p})a^{2} -
\frac{C^{2}_{\varphi}}{C}\rho_{\mathrm{CDM}}\Delta_{\mathrm{CDM}}a^{2}
&=& 0.
\end{eqnarray}
During the matter dominated era we can neglect the contribution
from radiation, and the above equation for the growth of the
overdensity reduces to
\begin{eqnarray}\label{eq:DeltaEvolution4}
\Delta''_{\mathrm{CDM}} + \tilde{H} \Delta'_{\mathrm{CDM}} &=& 4 \pi \tilde{G}
\rho_{\mathrm{CDM}}\Delta_{\mathrm{CDM}}a^{2},
\end{eqnarray}
where
\begin{eqnarray}
\tilde{G} \equiv \left(1 + {2C_\varphi^2 \over C} \right){\kappa \over 8\pi} &, & \tilde{H}  \equiv \frac{(aC)'}{a C}.
\end{eqnarray}
where $\kappa \equiv 8\pi G$ as before, and $2C_\varphi^2/C
\approx 2\gamma^2$ for our coupling function of $C$ as given by
Eq.~(\ref{eq:coupling_function}). Thus we could see that the
growth of cold dark matter density contrast is (on small scales)
scale-independent, which explains the matter power spectrum at
large $k$ values (small scales) in Fig.~\ref{fig:Figure5}. In
particular, note that for our choice of parameters we have $C \sim
1$ and  $\tilde{H} =  \frac{a'}{a} + \frac{C'}{C}  \sim
\frac{a'}{a} $, and so this equation is approximately the same as
that in the $\Lambda\mathrm{CDM}$ paradigm but with an effective
gravitational constant $\tilde{G} \sim (1+2\gamma^{2})G$. This
indicates that the parameter $\gamma$ characterizes the strength
of the scalar field fifth force relative to gravity; the larger
$\gamma$ is, the stronger the fifth force will \emph{ultimately}
be.

Meanwhile, from the derivation process of
Eq.~(\ref{eq:DeltaEvolution4}) we see that the assumption that
$k^{2}\gg
a^{2}\left(V_{\varphi\varphi}+\rho_{\mathrm{CDM}}C_{\varphi\varphi}\right)$
[cf.~Eq.~\ref{eq:sfpert_EOM2}] has been used. This is of course
true for our choices of $\mu$, but Eq.~(\ref{eq:V_phiphi}) tells
us that this is not necessarily the case for $\mu\ll1$, due to the
strong nonlinearity of $V(\varphi)$. Let's suppose now $k^{2}\ll
a^{2}\left(V_{\varphi\varphi}+\rho_{\mathrm{CDM}}C_{\varphi\varphi}\right)$
for some small enough value of $\mu$ (say $\mu=10^{-7}$), then
Eq.~(\ref{eq:sfpert_EOM2}) should be approximated as
\begin{eqnarray}\label{eq:sf_appox2}
(a^{2}V_{\varphi\varphi} +
a^{2}\rho_{\mathrm{CDM}}C_{\varphi\varphi})\xi +
k\varphi'\mathcal{Z} +
a^{2}C_{\varphi}\rho_{\mathrm{CDM}}\Delta_{\mathrm{CDM}} &\doteq&
0\nonumber
\end{eqnarray}
rather than as Eq.~(\ref{eq:sf_appox1}), so that in
Eq.~(\ref{eq:DeltaEvolution3}) the $k^{2}\xi$ term could be
neglected, implying that the scalar field simply has negligible
effects on the evolution of $\Delta_{\mathrm{CDM}}$. Thus we
conclude that the effect of the parameter $\mu$ is to control the
time when the fifth force becomes important as compared with
gravity; the smaller $\mu$ is, the later will this time be. So
with very strong chameleon effects (very nonlinear potentials) the
fifth force could be greatly suppressed all through the cosmic
history up to now, and it is therefore possible to get a cosmology
very close to the $\Lambda$CDM paradigm in every aspect.

Obviously, at the very early times the fifth force is also
suppressed even in cases of $\mu\sim\mathcal{O}(1)$, because in
this regime again we have $k^{2}\ll
a^{2}\left(V_{\varphi\varphi}+\rho_{\mathrm{CDM}}C_{\varphi\varphi}\right)$.
Indeed this is the reason why we could set the initial condition
as in Eq.~(\ref{eq:initcond}).

Now the matter power spectra as seen in Fig.~\ref{fig:Figure5}
could be well explained: when $\mu$ decreases, the time when the
fifth force could fully show its power becomes later and as a
result the growth of structure does not reach its maximum
potential -- this explains why the $\mu=0.1, 0.2, 0.3$ curves
increasingly get far away from the black one. But because all
these values of $\mu$ are large enough (compared with, say,
$\mu=10^{-7}$), the fifth force in all these cases has
\emph{almost} realized its \emph{full} power (which is
$2\gamma^{2}$ times stronger than gravity according to
Eq.~(\ref{eq:DeltaEvolution4})) -- this explains why all the
colored curves are fairly far from the black one. Finally,
Eq.~(\ref{eq:DeltaEvolution4}) explains why on small scales the
colored $P(k)$ curves are almost parallel to the black one.

Given the complexity of the scalar field behavior, especially in
regions with highly nonlinear matter distribution, we must say
that the linear analysis performed above is only qualitatively
correct and lacks the high precision to compare with various
cosmological data sets. In particular, it may well be that the
effect of the fifth force is not significant on linear scales
because the fifth force is not long-range enough, but much more
important on smaller scales which are beyond the linear regime.
Thus in the next section we shall consider the scalar field model
in the context of N-body cosmological simulations, and try to
study its effects in a more accurate manner.

\section{Nonlinear Structure Formation}

\label{sect:nonlin}

In \S~\ref{sect:lin} we have considered the background evolution
and linear large structure formation in the couple scalar field
model. For the former, the coupling between $\varphi$ and CDM
prevents $\varphi$ from rolling to infinity and instead tends to
keep $\varphi$ and $V(\varphi)$ constant; here we find that for a
value of $\mu\lesssim\mathcal{O}(0.1)$ this effect is significant
enough to make the background cosmology similar to that of
$\Lambda$CDM. For the latter, the CMB spectrum is also not very
sensitive to the value of $\mu$ or $\gamma$ such that
$\mu\lesssim\mathcal{O}(0.1)$ is difficult to be distinguished
from $\Lambda$CDM. These results suggest that large scale
observables are not particularly well suited in studying the new
features of this class of coupled scalar field models.

On smaller but still linear scales, we already see that the scalar
field coupling effectively increases the gravitational force by a
factor of $2\gamma^{2}$, leading to a significant increase in the
small scale power of $P(k)$ for gravitational-strength couplings
[$\gamma\sim\mathcal{O}(1)$]. One could of course argue that the
observed matter power spectrum is indeed for the luminous matter
only and cannot be applied to dark matter na\"{i}vely due to the
bias between the two. But the enhancement of dark matter
clustering in this model is abstract and can be tested by
observations such as weak lensing without much ambiguity. If this
is done, it is unlikely that there remains much space for such a
strongly coupled model.

However, two things must be taken into account before we arrive at
any definite conclusions about the fate of the model. First, on
small scales where the scalar field effect becomes important, the
distribution of matter is already beyond the linear perturbation
regime and in some regions could be very nonlinear. This means
that a linear analysis as presented in the above section is no
longer sufficient and we must consider the nonlinear effect, which
is most precisely taken in account by large N-body simulations.
This will be the very topic of this section. Second, as we argued
in \S~\ref{sect:lin}, the behavior of the model is not only
controlled by $\gamma$, but also by parameter $\mu$: for
$\mu\ll1$, the epoch when the scalar field fifth force starts to
realizes its full power as $2\gamma^{2}$ times of gravity could be
greatly postponed. It is then possible to prevent too powerful a
structure formation.

Like in the background cosmology, this has something to do with
nonlinear (chameleon) effects, but here things become much more
complicated. In a homogeneous universe, the spatial gradient of
the scalar field vanishes and the scalar field $\varphi$ takes the
same value anywhere, largely simplifying the analysis. For the
real universe with nonlinear matter distributions, the spatial
gradient of $\varphi$ is normally much larger than the time
derivative so that the configuration of the scalar field relies on
the underlying matter distribution sensitively in a nonlinear way,
and at the same time strongly affects the latter via the action of
fifth force. Quantifying such complex couplings also calls for the
use of N-body simulations.

In this paper we shall set up the basic framework of studying
coupled scalar field models using N-body technique, putting much
emphasis on the working mechanism and the fifth force effect. We
shall start with a derivation of all the relevant equations of
motion, describe how to integrate them into the numerical code and
present some preliminary results. Detailed analysis will be
postponed to companion papers. We will not provide an introduction
to general N-body simulation techniques here, and interested
readers are referred to the relevant literature
\cite{Bertschinger1998}.

\subsection{The Nonrelativistic Equations}

\label{subsect:noneqn}

Our first step is to simplify the relevant equations of motion in
the appropriate limit to get a set of equations which can be
directly applied to the numerical code.

We know that the existence of the scalar field and its coupling to
standard cold dark matter particles make the following changes to
the $\Lambda$CDM model: First, the energy momentum tensor has a
new piece of contribution from the scalar field; second, the
energy density of dark matter is multiplied by a function
$C(\varphi)$, which is because the coupling to scalar field
essentially renormalizes the mass of dark matter particles; third
and most important, dark matter particles will not follow
geodesics in their motions as in $\Lambda$CDM, rather, the total
force on them has a contribution from the scalar field.

These imply that the following things need to be modified or
added:
\begin{enumerate}
    \item The scalar field $\varphi$ equation of motion, which determines
    the value of the scalar field at any given time and position.
    \item The Poisson equation, which determines the gravitational
    potential (and thus gravity) at any given time and position,
    according to the local energy density and pressure, which
    include the contribution from the scalar field (as obtained from $\varphi$ equation of motion) now.
    \item The total force on the dark matter particles, which is
    determined by the spatial configuration of $\varphi$,
    just like gravity is determined by the spatial configuration
    of the gravitational potential.
\end{enumerate}
We shall describe these one by one now.

For the scalar field equation of motion, we denote $\bar{\varphi}$
as the background value of $\varphi$ and
$\delta\varphi\equiv\varphi-\bar{\varphi}$ as the scalar field
perturbation. Then Eq.~(\ref{eq:phiEOM}) could be rewritten as
\begin{eqnarray}
\ddot{\delta\varphi} + 3H\dot{\delta\varphi} +
\vec{\nabla}_{\mathbf{r}}^{2}\varphi +
V_{,\varphi}(\varphi)-V_{,\varphi}(\bar{\varphi})\nonumber\\ +
\rho_{\mathrm{CDM}}C_{,\varphi}(\varphi) -
\bar{\rho}_{\mathrm{CDM}}C_{,\varphi}(\bar{\varphi}) &=&
0\nonumber
\end{eqnarray}
by subtracting the corresponding background equation from it. Here
$\vec{\nabla}_{\mathbf{r}a}$ is the covariant spatial derivative
with respect to the physical coordinate $\mathbf{r}=a\mathbf{x}$
with $\mathbf{x}$ the conformal coordinate, and
$\vec{\nabla}^{2}_{\mathbf{r}}=\vec{\nabla}_{\mathbf{r}a}\vec{\nabla}_{\mathbf{r}}^{a}$.
$\vec{\nabla}_{\mathbf{r}a}$ is essentially the
$\hat{\nabla}_{a}$, but because here we are working in the weak
field limit we approximate it as $\vec{\nabla}_{\mathbf{r}}^{2} =
-\left(\partial^{2}_{r_{x}}+\partial^{2}_{r_{y}}+\partial^{2}_{r_{z}}\right)$
by assuming a flat background; the minus sign is because our
metric convention is $(+,-,-,-)$ instead of $(-,+,+,+)$. For the
simulation here we will also work in the quasi-static limit,
assuming that the spatial gradient is much larger than the time
derivative,
$|\vec{\nabla}_{\mathbf{r}}\varphi|\gg|\frac{\partial\varphi}{\partial
t}|$ (which will be justified below). Thus the above equation can
be further simplified as
\begin{eqnarray}\label{eq:WFphiEOM}
&&c^{2}\partial_{\mathbf{x}}^{2}(a\delta\varphi)\\ &=&
a^{3}\left[V_{,\varphi}(\varphi)-V_{,\varphi}(\bar{\varphi}) +
\rho_{\mathrm{CDM}}C_{,\varphi}(\varphi) -
\bar{\rho}_{\mathrm{CDM}}C_{,\varphi}(\bar{\varphi})\right],\nonumber
\end{eqnarray}
in which $\partial^{2}_{\mathbf{x}}=-\vec{\nabla}_{\mathbf{x}}^{2}
=
+\left(\partial^{2}_{x}+\partial^{2}_{y}+\partial^{2}_{z}\right)$
is with respect to the conformal coordinate $\mathbf{x}$ so that
$\vec{\nabla}_{\mathbf{x}}=a\vec{\nabla}_{\mathbf{r}}$, and we
have restored the factor $c^{2}$ in front of
$\vec{\nabla}_{\mathbf{x}}^{2}$ (the $\varphi$ here and in the
remaining of this paper is $c^{-2}$ times the $\varphi$ in the
original Lagrangian unless otherwise stated). Note that here $V$
and $\rho_{\mathrm{CDM}}$ both have the dimension of\emph{mass}
density rather than \emph{energy} density.

Next look at the Poisson equation, which is obtained from the
Einstein equation in weak-field and slow-motion limits. Here the
metric could be written as
\begin{eqnarray}
ds^{2} &=& (1+2\phi)dt^{2} - (1-2\psi)\delta_{ij}dr^{i}dr^{j}
\end{eqnarray}
from which we find that the time-time component of the Ricci
curvature tensor $R^{0}_{\ 0}=-\vec{\nabla}_{\mathbf{r}}^{2}\phi$,
and then the Einstein equation
$R_{ab}=\kappa\left(T_{ab}-\frac{1}{2}g_{ab}T\right)$ gives
\begin{eqnarray}\label{eq:EinsteinEqn}
R^{0}_{\ 0}\ =\ -\vec{\nabla}_{\mathbf{r}}^{2}\phi\ =\
\frac{\kappa}{2}(\rho_{\mathrm{TOT}}+3p_{\mathrm{TOT}})
\end{eqnarray}
where $\rho_{\mathrm{TOT}}$ and $p_{\mathrm{TOT}}$ are
respectively the total energy density and pressure. The quantity
$\vec{\nabla}_{\mathbf{r}}^{2}\phi$ can be expressed in terms of
the comoving coordinate $\mathbf{x}$ as
\begin{eqnarray}
\vec{\nabla}_{\mathbf{r}}^{2}\phi &=& \frac{1}{a^{2}}
\vec{\nabla}_{\mathbf{x}}^{2}\left(\frac{\Phi}{a} -
\frac{1}{2}a\ddot{a}\mathbf{x}^{2}\right)\nonumber\\
&=& \frac{1}{a^{3}}\vec{\nabla}_{\mathbf{x}}^{2}\Phi -
3\frac{\ddot{a}}{a}
\end{eqnarray}
where we have defined a new Newtonian potential
\begin{eqnarray}\label{eq:newphi}
\Phi &\equiv& a\phi + \frac{1}{2}a^{2}\ddot{a}\mathbf{x}^{2}
\end{eqnarray}
and used $\vec{\nabla}_{\mathbf{x}}^{2}\mathbf{x}^{2}=6$. Thus
\begin{eqnarray}
\vec{\nabla}_{\mathbf{x}}^{2}\Phi &=&
a^{3}\left(\vec{\nabla}_{\mathbf{r}}^{2}\phi +
3\frac{\ddot{a}}{a}\right)\\
&=& -
a^{3}\left[\frac{\kappa}{2}(\rho_{\mathrm{TOT}}+3p_{\mathrm{TOT}})
-
\frac{\kappa}{2}(\bar{\rho}_{\mathrm{TOT}}+3\bar{p}_{\mathrm{TOT}})\right]\nonumber
\end{eqnarray}
where in the second step we have used Eq.~(\ref{eq:EinsteinEqn})
and the Raychaudhrui equation, and an overbar means the background
value of a quantity. Because the energy momentum tensor for the
scalar field is given by Eq.~(\ref{eq:phiEMT}), it is easy to show
that $\rho^{\varphi}+3p^{\varphi} =
2\left[\dot{\varphi}^{2}-V(\varphi)\right]$ and so
\begin{eqnarray}
\vec{\nabla}_{\mathbf{x}}^{2}\Phi &=& -4\pi Ga^{3}
\left\{\rho_{\mathrm{CDM}}C(\varphi) +
2\left[\dot{\varphi}^{2}-V(\varphi)\right]\right\}\nonumber\\
&& +4\pi Ga^{3} \left\{\bar{\rho}_{\mathrm{CDM}}C(\bar{\varphi})+
2\left[\dot{\bar{\varphi}}^{2}-V(\bar{\varphi})\right]\right\}.\nonumber
\end{eqnarray}
Now in this equation $\dot{\varphi}^{2} - \dot{\bar{\varphi}}^{2}
= 2\dot{\varphi}\dot{\delta\varphi} + \dot{\delta\varphi}^{2} \ll
(\vec{\nabla}_{\mathbf{r}}\varphi)^{2}$ in the quasi-static limit
and so could be dropped safely. So we finally have
\begin{eqnarray}\label{eq:WFPoisson}
\partial_{\mathbf{x}}^{2}\Phi &=& 4\pi Ga^{3}
\left[\rho_{\mathrm{CDM}}C(\varphi)-\bar{\rho}_{\mathrm{CDM}}C(\bar{\varphi})\right]\nonumber\\
&&- 8\pi Ga^{3}\left[V(\varphi)-V(\bar{\varphi})\right].
\end{eqnarray}

Finally, for the equation of motion of the dark matter particle,
consider Eq.~(\ref{eq:DM_energy_conservation}). Using
Eqs.~(\ref{eq:DMEMT_particle}, \ref{eq:DMLagrangian2}), this can
be reduced to
\begin{eqnarray}\label{eq:DMEOM}
\ddot{x}^{a}_{0} + \Gamma^{a}_{bc}\dot{x}^{b}_{0}\dot{x}^{c}_{0}
&=&
\left(g^{ab}-u^{a}u^{b}\right)\frac{C_{\varphi}(\varphi)}{C(\varphi)}\nabla_{b}\varphi.
\end{eqnarray}
Obviously the left hand side is the conventional geodesic equation
and the right hand side is the new fifth force due to the coupling
to the scalar field. Before going on further, note that the fifth
force
$\frac{C_{\varphi}(\varphi)}{C(\varphi)}\hat{\nabla}_{a}\varphi=\hat{\nabla}_{a}\log
C(\varphi)$ is perpendicular to the 4-velocity $u_{a}$; this means
that the energy density of CDM will be conserved and only the
particle trajectories are modified as mentioned above. Now in the
non-relativistic limit the spatial components of
Eq.~(\ref{eq:DMEOM}) can be written as
\begin{eqnarray}\label{eq:DMEOM_physical}
\frac{d^{2}\mathbf{r}}{dt^{2}} &=& -\vec{\nabla}_{\mathbf{r}}\phi
-
\frac{C_{\varphi}(\varphi)}{C(\varphi)}\vec{\nabla}_{\mathbf{r}}\varphi
\end{eqnarray}
where $t$ is the physical time coordinate. If we instead use the
comoving coordinate $\mathbf{x}$, then this becomes
\begin{eqnarray}
\ddot{\mathbf{x}} + 2\frac{\dot{a}}{a}\dot{\mathbf{x}} &=&
-\frac{1}{a^{3}}\vec{\nabla}_{\mathbf{x}}\Phi  -
\frac{1}{a^{2}}\frac{C_{\varphi}(\varphi)}{C(\varphi)}
\vec{\nabla}_{\mathbf{x}}\varphi
\end{eqnarray}
where we have used Eq.~(\ref{eq:newphi}). The canonical momentum
conjugate to $\mathbf{x}$ is $\mathbf{p}=a^{2}\dot{\mathbf{x}}$ so
we have now
\begin{eqnarray}\label{eq:WFdxdtcomov}
\frac{d\mathbf{x}}{dt} &=& \frac{\mathbf{p}}{a^{2}},\\
\label{eq:WFdpdtcomov} \frac{d\mathbf{p}}{dt} &=&
-\frac{1}{a}\vec{\nabla}_{\mathbf{x}}\Phi -
\frac{C_{\varphi}(\varphi)}{C(\varphi)}\vec{\nabla}_{\mathbf{x}}\varphi.
\end{eqnarray}

Eqs.~(\ref{eq:WFphiEOM}, \ref{eq:WFPoisson}, \ref{eq:WFdxdtcomov},
\ref{eq:WFdpdtcomov}) will be used in the code to evaluate the
forces on the dark matter particles and evolve their positions and
momenta in time.

\subsection{Internal Units}

\label{subsect:intunit}

In our numerical simulation we use a modified version of MLAPM
(\cite{MLAPM}, see \ref{subsect:N-body}), and we must change our
above equations in accordance with the internal units used in that
code. Here we briefly summarize the main features.

MLAPM code uses the following internal units (with subscript
$_{c}$):
\begin{eqnarray}
\mathbf{x}_{c} &=& \mathbf{x}/B,\nonumber\\
\mathbf{p}_{c} &=& \mathbf{p}/(H_{0}B)\nonumber\\
t_{c} &=& tH_{0}\nonumber\\
\Phi_{c} &=& \Phi/(H_{0}B)^{2}\nonumber\\
\rho_{c} &=& \rho/\bar{\rho},
\end{eqnarray}
in which $B$ is the present size of the simulation box and $H_{0}$
is the present Hubble constant. Using these newly-defined
quantities, it is easy to check that Eqs.~(\ref{eq:WFdxdtcomov},
\ref{eq:WFdpdtcomov}, \ref{eq:WFPoisson}, \ref{eq:WFphiEOM}) could
be rewritten as
\begin{eqnarray}\label{eq:INTdxdtcomov}
\frac{d\mathbf{x}_{c}}{dt_{c}} &=& \frac{\mathbf{p}_{c}}{a^{2}},\\
\label{eq:INTdpdtcomov} \frac{d\mathbf{p}_{c}}{dt_{c}} &=&
-\frac{1}{a}\nabla\Phi_{c} -\frac{C_{,\varphi}}{C}c^{2}\nabla\varphi,\\
\label{eq:INTPoisson}\nabla^{2}\Phi_{c} &=&
\frac{3}{2}\Omega_{\mathrm{CDM}}\bar{C}
\left(\rho_{c}\frac{C}{\bar{C}}-1\right) -
\kappa\frac{V-\bar{V}}{H^{2}_{0}}a^{3},
\end{eqnarray}
and
\begin{eqnarray}\label{eq:INTphiEOM}
&&\frac{c^{2}}{\left(BH_{0}\right)^{2}}\nabla^{2}\left(a\varphi\right)\nonumber\\
&=&
\frac{3}{\kappa}\Omega_{\mathrm{CDM}}\bar{C}_{,\varphi}\left(\rho_{c}\frac{C_{,\varphi}}{\bar{C}_{,\varphi}}-1\right)
+ \frac{V_{,\varphi}-\bar{V}_{,\varphi}}{H^{2}_{0}}a^{3},
\end{eqnarray}
where $\Omega_{\mathrm{CDM}}$ is the present CDM fractional energy
density, we have again restored the factor $c^{2}$ and again the
$\varphi$ is $c^{-2}$ times the $\varphi$ in the original
Lagrangian. Also note that from here on we shall use
$\nabla\equiv\vec{\partial}_{\mathbf{x}_{c}},
\nabla^{2}\equiv\vec{\partial}_{\mathbf{x}_{c}}\cdot\vec{\partial}_{\mathbf{x}_{c}}$
unless otherwise stated, for simplicity.

We also define
\begin{eqnarray}
\chi &\equiv& \sqrt{\kappa}\varphi,\\
u & \equiv & \ln(e^\chi -1) \\
\Omega_{V_{0}} &\equiv& \frac{\kappa V_{0}}{3H_{0}^{2}},
\end{eqnarray}
to be used below.

Making discretized version of the above equations for N-body
simulations is non-trivial task. For example, the use of variable
$u$ instead of $\varphi$ (see below) helps to prevent $\varphi<0$,
which is unphysical, but numerically possible due to
discretization. We refer the interested readers to
Appendix~\ref{appen:discret} to the whole treatment, with which we
can now proceed to do N-body runs.

\subsection{The N-Body Code}

\label{subsect:N-body}

The full name of MLAPM is Multi-Level Adaptive Particle Mesh code.
As the name has suggested, this code uses multilevel grids
\cite{Brandt1977, Press1992, Briggs2000} to accelerate the
convergence of the (nonlinear) Gauss-Seidel relaxation method
\cite{Press1992} in solving boundary value partial differential
equations. But more than this, the code is also adaptive, always
refining the grid in regions where the mass/particle density
exceeds a certain threshold. Each refinement level form a finer
grid which the particles will be then (re)linked onto and where
the field equations will be solved (with a smaller time step).
Thus MLAPM has two kinds of grids: the domain grid which is fixed
at the beginning of a simulation, and refined grids which are
generated according to the particle distribution and which are
destroyed after a complete time step.

One benefit of such a setup is that in low density regions where
the resolution requirement is not high, less time steps are
needed, while the majority of computing sources could be used in
those few high density regions where high resolution is needed to
ensure precision.

Some technical issues must be taken care of however. For example,
once a refined grid is created, the particles in that region will
be linked onto it and densities on it are calculated, then the
coarse-grid values of the gravitational potential are interpolated
to obtain the corresponding values on the finer grid. When the
Gauss-Seidel iteration is performed on refined grids, the
gravitational potential on the boundary nodes are kept constant
and only those on the interior nodes are updated according to
Eq.~(\ref{eq:GS}): just to ensure consistency between coarse and
refined grids. This point is also important in the scalar field
simulation because, like the gravitational potential, the scalar
field value is also evaluated on and communicated between
multi-grids (note in particular that different boundary conditions
lead to different solutions to the scalar field equation of
motion).

In our simulation the domain grid (the finest grid that is not a
refined grid) has $128^{3}$ nodes, and there are a ladder of
coarser grids with $64^3$, $32^3$, $16^3$, $8^3$, $4^3$ nodes
respectively. These grids are used for the multi-grid acceleration
of convergence: for the Gauss-Seidel relaxation method, the
convergence rate is high upon the first several iterations, but
quickly becomes very slow then; this is because the convergence is
only efficient for the high frequency (short-range) Fourier modes,
while for low frequency (long-range) modes more iterations just do
not help much. To accelerate the solution process, one then
switches to the next coarser grid for which the low frequency
modes of the finer grid are actually high frequency ones and thus
converge fast. The MLAPM solver adopts the self-adaptive scheme:
if convergence is achieved on a grid, then interpolate the
relevant quantities back to the finer grid (provided that the
latter is not on the refinements) and solve the equation there
again; if convergence becomes slow on a grid, then go to the next
coarser grid. This way it goes indefinitely except when converged
solution on the domain grid is obtained or when one arrives at the
coarsest grid (normally with $2^3$ nodes) on which the equations
can be solved exactly using other techniques. For our scalar field
model, the equations are difficult to solve anyway, and so we
truncate the coarser-grid series at the $4^3$-node one, on which
we simply iterate until convergence is achieved.

For the refined grids the method is different: here one just
iterate Eq.~(\ref{eq:GS}) until convergence, without resorting to
coarser grids for acceleration.

As is normal in the Gauss-Seidel relaxation method, convergence is
deemed to be achieved when the numerical solution $u^{k}_{n}$
after $n$ iterations on grid $k$ satisfies that the norm
$\Vert\cdot\Vert$ (mean or maximum value on a grid) of the
residual
\begin{eqnarray}
e^{k} &=& L^{k}(u^{k}_{n}) - f_{k},
\end{eqnarray}
is smaller than the norm of the truncation error
\begin{eqnarray}
\tau^{k} &=& L^{k-1}(\mathcal{R}u^{k}_{n}) -
\mathcal{R}\left[L^{k}(u^{k}_{n})\right]
\end{eqnarray}
by a certain amount. Note here $L^{k}$ is the discretization of
the differential operator Eq.~(\ref{eq:diffop}) on grid $k$ and
$L^{k-1}$ a similar discretization on grid $k-1$, $f_k$ is the
source term, $\mathcal{R}$ is the restriction operator to
interpolate values from the grid $k$ to the grid $k-1$. In the
modified code we have used the full-weighting restriction for
$\mathcal{R}$. Correspondingly there is a prolongation operator
$\mathcal{P}$ to obtain values from grid $k-1$ to grid $k$, and we
use a bilinear interpolation for it. For more details see
\cite{MLAPM}.

MLAPM calculates the gravitational forces on particles by centered
difference of the potential $\Phi$ and propagate the forces to
locations of particles by the so-called triangular-shaped-cloud
(TSC) scheme to ensure momentum conservation on all grids. The TSC
scheme is also used in the density assignment given the particle
distribution.

The main modifications to the MLAPM code for our model are:
\begin{enumerate}
    \item We have added a parallel solver for the scalar field
    based on Eq.~(\ref{eq:u_phi_EOM}). The solver uses a similar
    nonlinear Gauss-Seidel method and same criterion for
    convergence as the Poisson solver.
    \item The solved value of $u$ is then used to calculate local
    mass density and thus the source term for the Poisson
    equation, which is solved using fast Fourier transform.
    \item The fifth force is obtained by differentiating the $u$
    just like the calculation of gravity.
    \item The momenta and positions of particles are then updated
    taking in account of both gravity and the fifth force.
\end{enumerate}
There are a lot of additions and modifications to ensure smooth
interface and the newly added data structures. For the output, as
there are multilevel grids all of which host particles, the
composite grid is inhomogeneous and thus we choose to output the
positions, momenta of the particles, plus the gravity, fifth force
and scalar field value \emph{at the positions} of these particles.
We can of course easily read these data into the code, calculate
the corresponding quantities on each grid and output them if
needed.

\subsection{Preliminary Numerical Results}

\label{subsect:numresult}

In this subsection we shall present some preliminary results of
several runs and give a sense about the qualitative behaviors of
the coupled scalar field model. Also we will not make detailed and
quantitative analysis in this paper, which will be shown in
forthcoming papers.

We have performed 8 runs of the modified code with parameters
$\gamma=0.5, 1$ and $\mu=10^{-4}, 10^{-5}, 10^{-6}, 10^{-7}$
respectively. For all these runs there are $128^{3}$ dark matter
particles, the simulation box has a size $B =
64h^{-1}~\mathrm{Mpc}$ in which $h=H_{0}/(100~\mathrm{km/s/Mpc})$
and $128$ domain grid cells in each direction. We assume a
$\Lambda$CDM background cosmology which is a very good
approximation for $\mu\ll1$ as we mentioned in \S~\ref{sect:lin};
the current fractional energy densities of dark matter and dark
energy are $\Omega_{\mathrm{CDM}}=0.28$ and
$\Omega_{\Lambda}=0.72$ (note that this is a dark-matter-only
simulation and baryons will be added in a later work to study the
bias effect caused by the dark matter coupling).

Given these parameters, the mass resolution of the simulation is
$9.71\times10^{9}~M_{\bigodot}$ with $M_{\bigodot}$ the solar
mass. The spatial resolution is $\sim23.44h^{-1}~\mathrm{kpc}$ on
the finest refined grids and $0.5h^{-1}~\mathrm{Mpc}$ on the
domain grid. The high resolution in high density regions is
actually necessary to ensure precision in those regions because
the fifth force is generally short-ranged there.

All the simulations start at redshift $z=49$. In principle,
modified initial conditions (initial displacements and velocities
of particles which is obtained given a linear matter power
spectrum) need to be generated for the coupled scalar field model,
because the Zel'dovich approximation \cite{Zeldovich,
Efstathiou1985} is also affected by the scalar field coupling. In
practice, however, we find that the effect on the linear matter
power spectrum is negligible ($\lesssim\mathcal{O}(10^{-4})$) for
our choices of parameters $\gamma, \mu$. Thus we simply use the
$\Lambda$CDM initial displacements/velocities for the particles in
these simulations, which are generated using GRAFIC \cite{GRAFIC}
again with $\Omega_{\mathrm{CDM}}=0.28$ and
$\Omega_{\Lambda}=0.72$. $\sigma_{8}=0.88$ at present day.

\begin{figure*}[tbp]
\centering
\includegraphics[scale=0.85]{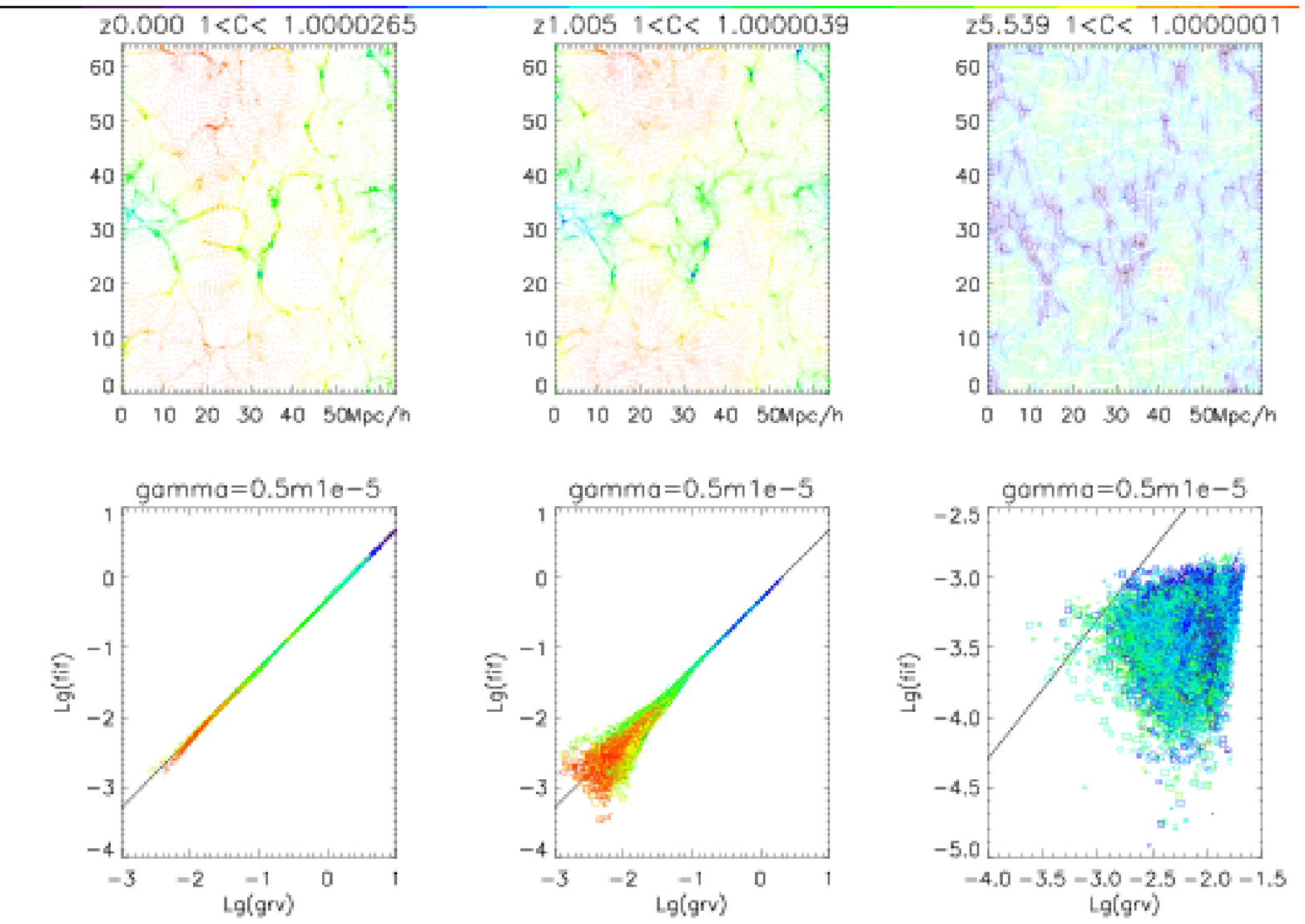}
\includegraphics[scale=0.85]{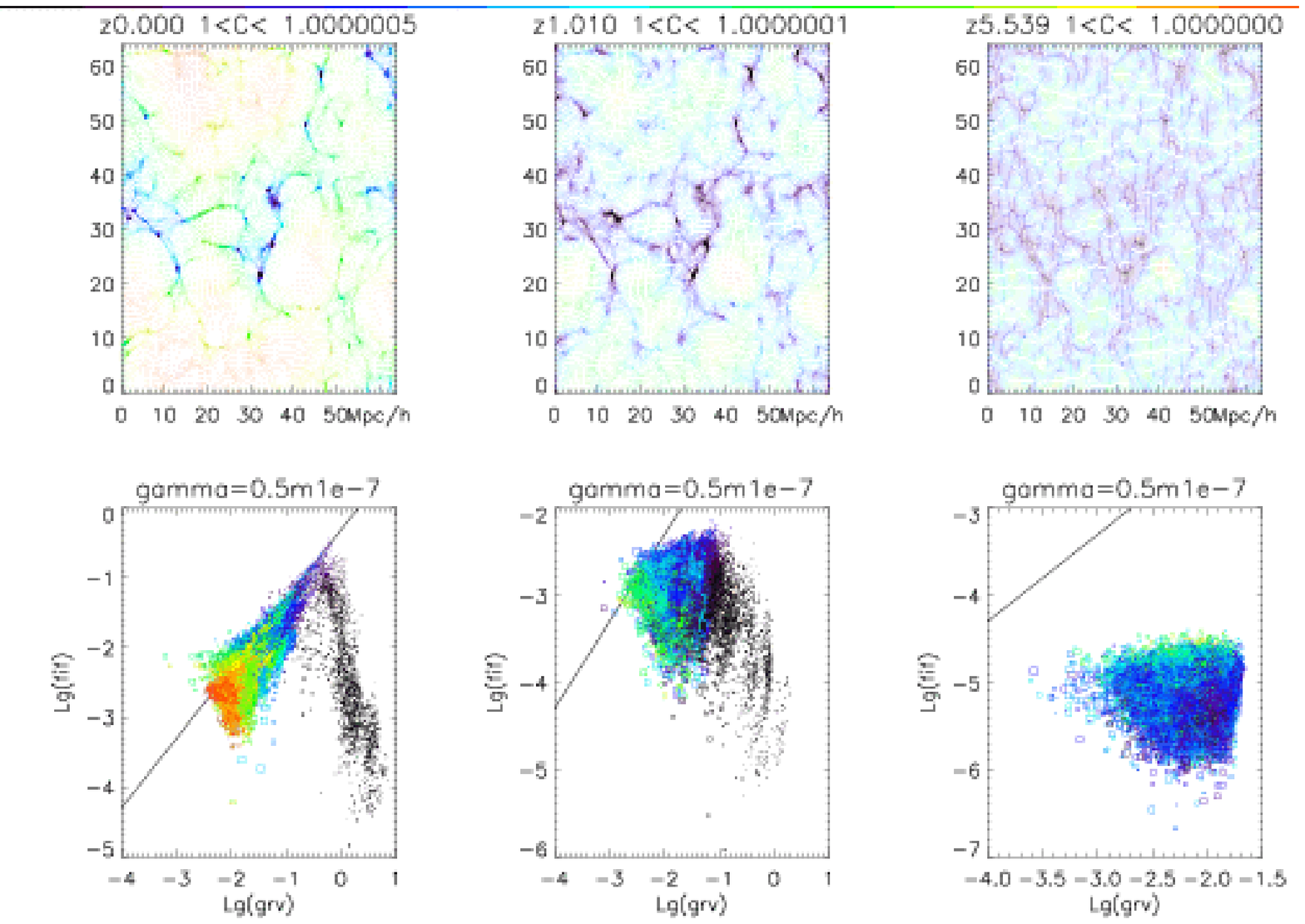}
\caption{(Color Online) The simulation results for the
$\gamma=0.5$ and $\mu=10^{-5}, 10^{-7}$ runs at redshifts $z=0$,
$z=1$, and $z=5.5$. Shown are the particle spatial yz distribution
in a slab of $x=31.5-32.5$Mpc/h (upper panel) and the lg-lg
diagram of the fifth force vs. gravity in this slab (lower panel).
See text for a detailed description. Each particle is a symbol
with its color denoting the the value of the effective mass
$C(\varphi)$, whose minimum/maximimu are given on top of each
panel, and correspond to the two ends of the color scale shown on
the top. The size of square symbols in lower panels are
proportional to the mis-alignment angle between the fifth force
and the gravity on a particle; the biggest squares correspond to
anti-alignment, and particles with well-aligned forces are shown
as dots.  A line showing a fifth-to-gravity ratio of $2\gamma^2$
is also drawn to show the (lack of) correlations of the two
forces.  Note that $C(\varphi)$ is generally the biggest in voids
where the forces are the weakest, poorly aligned and
less-correlated.} \label{fig:Figurex1}
\end{figure*}

\begin{figure*}[tbp]
\centering
\includegraphics[scale=0.85]{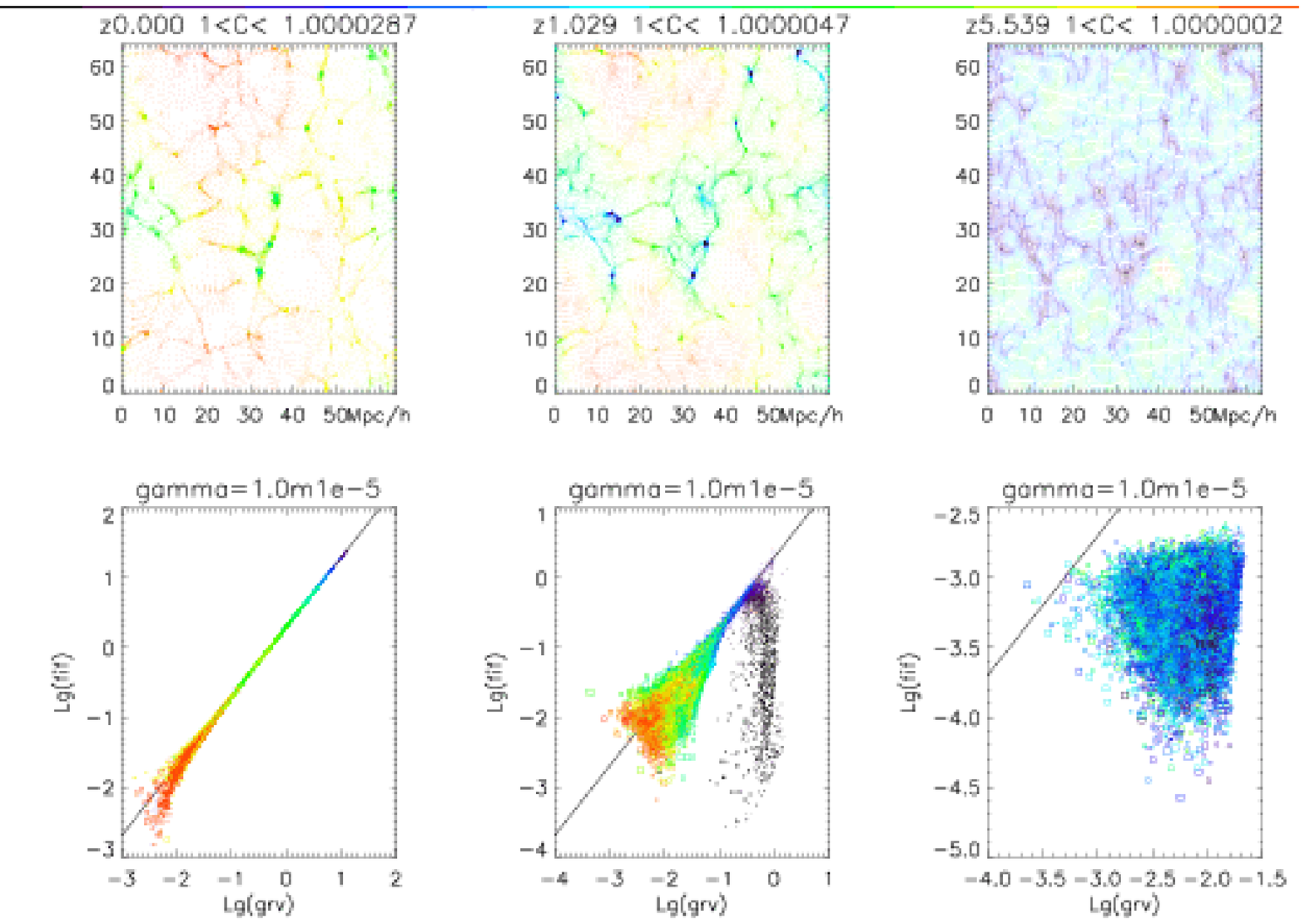}
\includegraphics[scale=0.85]{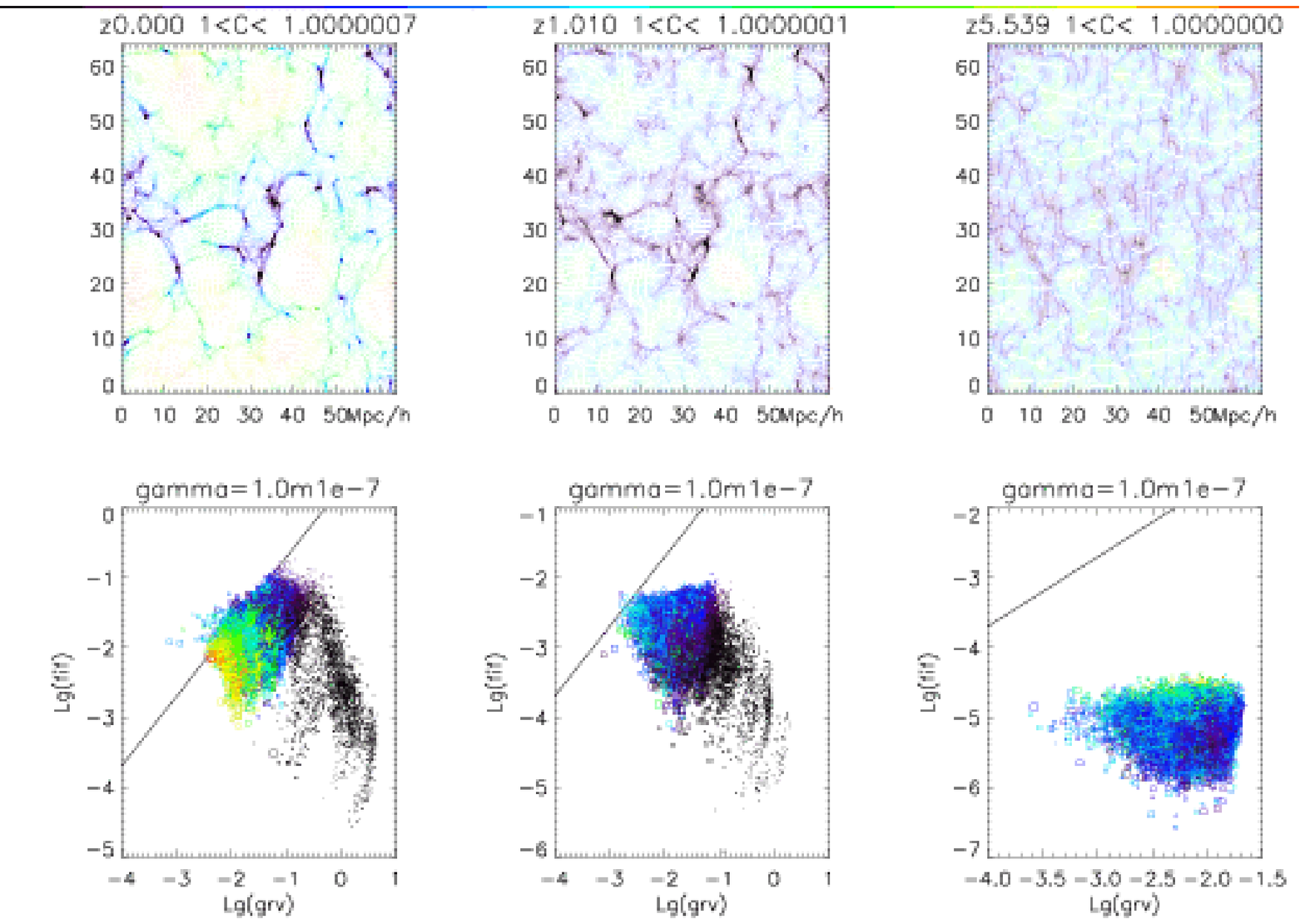}
\caption{(Color Online) The same as Fig.~\ref{fig:Figurex1} but
for the $\gamma=1.0$ and $\mu=10^{-5}, 10^{-7}$ runs.}
\label{fig:Figurex2}
\end{figure*}

In Figs.~\ref{fig:Figurex1}, \ref{fig:Figurex2} we have shown the
results for the runs with $\gamma=0.5, 1.0$ and $\mu=10^{-5},
\mu^{-7}$. These choices of $\mu$ are such that for $\mu =
10^{-7}$ (the lower two rows) the chameleon effect is pretty
strong while for $\mu = 10^{-5}$ (the upper two rows) it is much
weaker; the choices of $\gamma$ are to see the effects of
different full strengths of the fifth force. The three panels in
the first and third rows display the particle distributions at
three output redshifts $z\doteq5.5,~1.0,~0.0$ from right to left,
and the three panels in the second and fourth rows show the
correlation between the fifth force and gravity at these
redshifts.

For clearness we have only plotted a thin slice (along the $x$
direction) of the full 3-dimensional particle distribution. First
let's have a look at the first and third rows. Because we have
output the scalar field value at the position of each particle
together with other parameters of the particle, we also include
this information in the plots. On top of each panel the range of
the value $C(\varphi)$ at the positions of all the shown particles
is shown, and color is used to illustrate the amplitude of
$C(\varphi)$ (going from black to red from the minimum to the
maximum value of $C(\varphi)$). Note that
$C(\varphi)=\exp(\gamma\sqrt{\kappa}\varphi)\doteq1+\gamma\sqrt{\kappa}\varphi$
for $\sqrt{\kappa}\varphi\ll1$, thus the color also indicates the
value of the scalar field indirectly. Also bear in mind that the
same color may denote different values of $C(\varphi)$ at
different redshifts. The scalar field or $C(\varphi)$ grows with
time in general.

Next look at the second and fourth rows, which display the
logarithmic of the magnitude of the fifth force versus that of
gravity. The color here has the same meaning as above.
Furthermore, each point (particle) now is engaged with a square
box centered on it, which denotes the size of the angle (from $0$
to $\pi$) between the two forces; the size of the box increases
linearly with the angle, from a minimum size $0$ to a maximum size
comparable to the largest box size shown in the entire figure. As
we have mentioned above, the strength of the fifth force, if not
suppressed by the chameleon mechanism, is $2\gamma^{2}$ times that
of gravity, so we also plot the functions
\begin{eqnarray}\label{eq:FGcurve}
\lg F &=& \lg G + \lg(2\gamma^{2}),
\end{eqnarray}
where $F, G$ are respectively the magnitudes of the fifth force
and gravity, as the black solid lines in these figure, to compare
with the simulation data.

We can understand these results qualitatively as follows by taking
Fig.~\ref{fig:Figurex1} as example. The chameleon effect is
generally stronger at earlier times when matter density is high
and the scalar field value is small. As is shown in the panels of
the first row, at redshift $z=5.5$ there is a strong contrast of
the value $C(\varphi)$ in high density (\emph{blue}) regions
(clusters hereafter) compared to $C(\varphi)$ in the low density
(\emph{green, yellow and red}) regions (voids hereafter), which is
a direct reflection of the nonlinearity in the scalar field. As
time evolves and the background value of the scalar field
increases, at redshift $z=1.0$ the chameleon effect gets
suppressed; this is manifested by the facts that (1) in the
regions of small clusters the scalar field values are no longer
significantly different from the background value, both of which
are yellow and orange-colored, (2) even in the largest clusters
the contrast between the scalar field values inside and outside
(green/light blue versus yellow/orange) is not so strong compared
with the result at redshift $z=5.5$ (purple/dark blue versus
green/yellow). These show that the size of nonlinear regions is
shrinking and the thin shells in the clusters are thickening,
together leading to less nonlinear behaviors of the scalar field.
At redshift $z=0.0$ this tendency just becomes more obvious,
leaving only a small portion of the space with chameleon effect
and thin shells.

The strength of the chameleon effect is determined by several
factors. In principle, the larger the effective mass of scalar
field ($m_{eff}$) at a position is, the shorter-ranged the fifth
force is \footnotemark[1] and the less sensitive the scalar field
value at this position will be to the matter distribution around
it: this in turn corresponds to a stronger chameleon effect
because the scalar field value is mainly determined by the local
matter density. On the other hand, $m_{eff}$ depends on $\mu$
(smaller $\mu$ implies larger $m_{eff}$ within a given cluster and
thus stronger chameleon effect), $\gamma$ and local
$\rho_{\mathrm{CDM}}$ (larger values for these two parameters also
imply larger $m_{eff}$ and stronger chameleon effect) and also the
background value of the scalar field (this is because the interior
solution of the scalar field inside a cluster should only be
solved given the boundary conditions outside the cluster, due to
the differential nature of the scalar field EOM. Of course for
very small $\mu$ and/or very large $\gamma$, the scalar field EOM
indeed behaves as an algebraic equation and then the influence of
the background value of the scalar field is not important, but for
our choices of $\mu, \gamma$ that influence is significant). These
analysis agree well with what we have seen in the scalar field
configuration from the above N-body simulation results (first and
third rows of Fig.~\ref{fig:Figurex1}). Also note that the fifth
force is much weaker in regions where chameleon effect is strong,
because a particle there can only feel the fifth force from those
particles that are very close to it: in this sense we say that the
chameleon effect could suppress the fifth force.

\footnotetext[1]{Note that strictly speaking the fifth force
between two particles in this model depends on the detailed matter
distribution between these particles, and it is not very accurate
to simply relate it to a certain scalar field mass $m_{eff}$.
However, the concept of a fifth force whose range is $\sim
m_{eff}^{-1}$ is qualitatively correct and can help understand the
situation more intuitively.}

A comparison between gravity and the fifth force can illustrate
this more clearly. Remember that the chameleon effect could
strongly suppress the fifth force. At low redshift ($z=0.0$) the
chameleon effect is not significant and so the fifth force is not
suppressed; in this case we find that there is a strong
correlation between both the magnitudes and directions of these
two forces, and the simulation results agree with the prediction
Eq.~(\ref{eq:FGcurve}) to very high degrees. Both forces are
stronger in clusters and weaker in voids as expected. Going
backwards in time to the redshift $z=1.0$, there is still a good
correlation but the simulation results begin to scatter over the
predicted line Eq.~(\ref{eq:FGcurve}) in the void regions due to
the chameleon effect (the scalar field potential term in
Eq.~(\ref{eq:INTphiEOM}) becomes comparable with the matter
coupling term). Finally, at very early times ($z=5.5$) the scalar
field value becomes so small that the potential term in
Eq.~(\ref{eq:INTphiEOM}) is indeed much more important than the
matter coupling term so that the latter can be neglected: we then
have a complete mismatch between the numerical results and the
prediction Eq.~(\ref{eq:FGcurve}) as almost all data points are
significantly below the black solid line.

The above mismatch in the strongly chameleon regime can be
understood schematically as follows:
\begin{enumerate}
    \item When there is no potential for the scalar field but just a matter
    coupling, the fifth force is indeed long ranged and can probe the
    same region as gravity does. Then a comparison of
    Eqs.~(\ref{eq:INTdpdtcomov}, \ref{eq:INTPoisson},
    \ref{eq:INTphiEOM}) shows that the fifth force is exactly
    $2\gamma^{2}$ times of gravity. This is indeed what we have
    observed for $z=0.0$ when the potential term in
    Eq.~(\ref{eq:INTphiEOM}) is negligible.
    \item When the scalar field has
    a potential, it acquires an effective mass which is well-known to
    be inversely proportional to the range of the scalar fifth force.
    As a result the fifth force is no longer as long range as
    gravity. Meanwhile, Eq.~(\ref{eq:INTdpdtcomov}) makes it clear that the scalar
    field $\varphi$ (or equivalently $\ln[C(\varphi)]$) acts as a
    potential for the fifth force, and Eq.~(\ref{eq:INTphiEOM})
    shows that this potential depends on the underlying matter
    distribution in a different (nonlinear) way from what the gravitational potential
    $\Phi$ does.

    So we could see that there will be differences between both the ranges and the magnitudes of gravity and fifth
    force. In high density regions these differences will be dominated
    over by the competing effects that (\emph{i}) the majority part of the (either
    gravitational or fifth) force on a particle is contributed by
    nearby particles and there are so many particles nearby that
    contribution from distant particles is negligible, (\emph{ii})
    in Eq.~(\ref{eq:INTphiEOM}) the potential term is much less
    important than the matter coupling term, together making the
    fifth force behavior similar to that of gravity again. In the void regions
    the number of nearby particles is small so that contribution from
    distant particles should be taken into account, and the matter
    coupling term in Eq.~(\ref{eq:INTphiEOM}) is much smaller, then the
    difference between gravity and the fifth force becomes
    manifesting. These effects can be observed in the $z=1.0$ panel.
    \item At even higher redshifts the potential of the scalar field makes
    its effective mass very large and thus its range very short compared
    with that of gravity. Meanwhile Eq.~(\ref{eq:INTphiEOM}) becomes
    very nonlinear due to the smallness of the scalar field value, and thus
    the fifth force potential depends on matter distribution very differently
    from the gravitational potential.

    As a result, the fifth force reflects the
    matter distribution in a very small region around a given
    particle, while gravity probes that in a much larger region.
    Even in that small region where both forces exist, their
    magnitude can generally be very different. So it is not surprising that the two forces look so
    different as in the $z=5.5$ panel. Note that here the fifth
    force is also much weaker than gravity because its strength is
    suppressed by the smallness of the scalar field value.
\end{enumerate}

Result for the $\mu=10^{-7}$ case (the lower two rows) is
qualitatively the same as that of the $\mu=10^{-5}$ case above,
but here because $\mu$ is much smaller, so the chameleon effect
exists until much more recent than in the $\mu=10^{-5}$ case.
Indeed, even at low redshifts $z=1.0$ and $0.0$ there is still
strong contrast between the scalar field values inside and outside
the clusters (purple/dark blue versus yellow/orange). The
correlation between the forces also follow our above analysis, but
here there are some new features. The first feature is that even
today the fitting to Eq.~(\ref{eq:FGcurve}) is far from perfect;
this is easy to understand, because the scalar field potential is
so nonlinear that the scalar field potential term in
Eq.~(\ref{eq:INTphiEOM}) is important up to now. The second
feature is that in the $z=0.0$ (also $z=1.0$) panel we could find
that in high density regions the fifth force does not obey
Eq.~(\ref{eq:FGcurve}) as in the $\mu=10^{-5}$ case, but becomes
much smaller than gravity -- actually, the stronger gravity is,
the weaker the fifth force will be! This is again due to the
strong chameleon effect in the clusters, which makes the scalar
field value very small and thus suppresses the fifth force there.
Note that this feature is desirable because if we also couple the
scalar field to baryonic matter then we definitely want the fifth
force to be suppressed to evade solar system tests.

If we choose $\gamma=1.0$ as in Fig.~\ref{fig:Figurex2}, then all
the qualitative results we have obtained above should still apply.
But the stronger coupling between matter and the scalar field
enhances the chameleon effect. This implies that a coupling whose
strength is significantly larger than that of gravity could
probably produce the correct amount of large scale structure as
observed by suppressing the fifth force on all cosmological epochs
and scales of interests to us.

The analysis above clearly shows the complexity of the fifth force
and its possible effects on the nonlinear structure formation.
Though in certain (no chameleon) limits the fifth force is greatly
simplified and one can assume a modified gravitational constant in
the N-body simulations as an approximation, this is evidently not
the case if the scalar field potential is too much nonlinear. Of
course, we have no \emph{a priori} knowledge about when the
chameleon effect becomes important, and full numerical simulations
like the one presented here are therefore necessary to precisely
study the effects of (coupled) scalar fields in the large scale
structure. In forthcoming works we shall analyze the nonlinear
matter power spectrum, halo profile, scalar field configuration
within clusters, as well as the development and disappearance of
thin shells in a quantitative manner.

One may wonder if a more complete simulation should keep the time
derivatives of the scalar field in Eq.~(\ref{eq:WFphiEOM}). This
is certainly true, yet these terms indeed have negligible effect
\cite{Oyaizu2008}, which could be understood in the following way:
when there is no (or very weak) chameleon effect, the scalar field
potential term in Eq.~(\ref{eq:INTphiEOM}) is negligible and so
this equation has the same form as the Poisson equation
Eq.~(\ref{eq:INTPoisson}); as a result the quasi-static
approximation works as well for the scalar field as for the
gravitational potential (which is what any N-body code relies on).
On the other hand, if there is strong chameleon effect, the scalar
field value tends to be much smaller which means that the time
derivative of the scalar field also gets smaller; and at the same
time the spatial gradient of the scalar field becomes larger:
these together indicate that the quasi-static approximation should
be good here too.

\section{Conclusion}

\label{sect:disc}

To conclude, in this paper we have presented the general
frameworks to study the linear and nonlinear structure formations
in coupled scalar field models, and given some preliminary
numerical results for both in the context of a specific coupling
function and scalar field potential.

For the linear large scale structure, we write down the perturbed
field equations using the $3+1$ decomposition, which can be
directly applied into numerical Boltzmann codes such as CAMB to
generate the CMB and matter power spectra. For the chosen coupling
function with parameter $\gamma$ and potential with parameter
$\mu$, we find that $\gamma$ roughly controls the strength of the
fifth force (which is due to the propagation of the scalar field)
while $\mu$ controls how much the effect of the fifth force is
suppressed by the chameleon mechanism. With
$\mu\lesssim\mathcal{O}(0.1)$ the chameleon effect makes the model
behave like $\Lambda$CDM on very large scales, but this is far
from enough to significantly decrease the effects of the fifth
force on the small scale density perturbation growth. On those
small scales, however, nonlinearity becomes an important issue,
which leads us to the N-body simulation in \S~\ref{sect:nonlin}.

Previous N-body simulations with scalar fields are generally
simplified by certain approximations such as treating the scalar
coupling effect as a simple change of gravitational constant $G$,
or assuming a Yukawa-type force with a certain range. These
approximations do not work well in the present model, and so here
we set up the formula needed for a more precise simulation,
putting much emphasis on the calculation of the fifth force and
its action on particles. The equations derived and the algorithm
described in \S~\ref{sect:nonlin} are general enough and should be
directly applicable to other coupled scalar field models.

We integrate the equations into a modified version of the N-body
code MLAPM and performed several runs with different combinations
of $\gamma, \mu$. Some results are displayed in
Figs.~\ref{fig:Figurex1} and \ref{fig:Figurex2}. It is confirmed
that when the chameleon effect is not important, the fifth force
is parallel to gravity and is $2\gamma^{2}$ times stronger,
meaning that simply using a different gravitational constant which
is $2\gamma^{2}+1$ times as large as the bare one in the
simulation should be an acceptable approximation. There are some
caveats however. For one thing, the correlation between gravity
and the fifth force is only good for a portion of the parameter
space $(\gamma, \mu)$, and for $\mu\ll1$ the potential is so
nonlinear as to destroy this correlation. More importantly, even
the correlation is perfect now it is possible that at earlier
times the nonlinear effects modify it dramatically
(cf.~Fig.~\ref{fig:Figurex2} with $\mu=10^{-5}$ at $z=0.0, 1.0$).
This means that the above-mentioned approximation is unlikely to
be correct consistently and full simulations as the ones in this
paper are needed.

All in all, from the results we can spot the trend that, with the
same value of $\mu$ increasing $\gamma$ simply enhances the power
of structure growth, while with the same value of $\gamma$
decreasing $\mu$ has the effects of reducing that power. Also, for
small $\mu$ the configuration of the scalar field becomes very
nonlinear and highly sensitive to the underlying matter
distribution, while for large $\mu$ this becomes much more smooth
and stiff. All these observations (and others as explained in
\S~\ref{subsect:numresult}) agree with the chameleon analysis. Our
simulations also provides a test bed for scalar field theories.
Among the applications, one could look for high-speed encounters
of halos to see the probability of generating a
Bullet-cluster-like encounters \cite{Springel2007, Claudio2009}.
Since this paper is only served to set up the general framework of
linear and nonlinear simulations, these applications of the
simulation results, as well as the matter power spectrum and
scalar field profile/evolution, will be presented in forthcoming
works.

\begin{acknowledgments}
The authors thank Alexander Knebe, Claudio Llinares and Xufen Wu
for their helps in technical problems about the N-body code and
its implementation, John Barrow, Carsten van der Bruck, Anne
Davis, George Efstathiou, David F. Mota and Douglas Shaw for
encouragements to finish this work and discussions in the process,
and Luca Amendola for listening about the work and helpful
information at earlier stages. B.~Li acknowledges supports from
Overseas Research Studentship, Cambridge Overseas Trust, DAMTP and
Queens' College. We are also indebted to the HPC-Europa
Transnational Access Visit programme for its support and the
Lorentz Center and Leiden Observatory for hospitality when part of
this work is undertaken. The N-body simulations are performed on
the SARA supercomputer in the Netherlands.
\end{acknowledgments}

\bigskip


\appendix

\section{The Perturbation Equations}

\label{appen:perteqn}

Consider the decomposition of the stress energy tensor $T_{ab}$
\begin{eqnarray}
T_{ab} &=& \pi_{ab} + 2q_{(a}u_{b)} + \rho u_{a}u_{b} - ph_{ab}
\end{eqnarray}
where $u_{a}$ is the 4-velocity of an observer with respect to
which $3+1$ space-time splitting is made, $h_{ab} = g_{ab} -
u_{a}u_{b}$ the projection tensor used to obtain covariant tensors
perpendicular to $u_{a}$. Here $\pi_{ab}$ is the projected
symmetric tracefree anisotropic stress, $q$ is the vector heat
flux and $\rho$ and $p$ respectively the energy density and
isotropic pressure. These quantities could be obtained from
$T_{ab}$ through the relations
\begin{eqnarray}
\rho &=& T_{ab}u^{a}u^{b},\nonumber\\
p &=& -\frac{1}{3}h^{ab}T_{ab},\nonumber\\
q_{a} &=& h^{d}_{a}u^{c}T_{cd},\nonumber\\
\pi_{ab} &=& h^{c}_{a}h^{d}_{b}T_{cd} + ph_{ab}.
\end{eqnarray}
Based on these, the components of the stress energy tensor (up to
first order) in this model is summarized in the following table:
\begin{table}[htbp]
\caption{\label{tab:table3} \emph{The decomposition of energy
momentum tensor in the $3+1$ formalism.}} \label{table}
\end{table}
\begin{center}
\begin{tabular}{ccccc}
\hline\hline
Matter & $\rho$ & $p$ & $q_{a}$ & $\pi_{ab}$ \\
\hline $\gamma$ & $\rho_{\gamma}$ & $\frac{1}{3}\rho_{\gamma}$ & $q_{\gamma a}$ & $\pi_{\gamma ab}$\\
$\nu$ & $\rho_{\nu}$ & $\frac{1}{3}\rho_{\nu}$ & $q_{\nu a}$ & $\pi_{\nu ab}$\\
Baryons & $\rho_{\mathrm{B}}$ & $p_{\mathrm{B}}$ & $q_{\mathrm{B}a}$ & $0$\\
CDM & $\rho_{\mathrm{CDM}}$ & $0$ & $q_{\mathrm{CDM}a}$ & $0$\\
Coupled CDM & $C(\varphi)\rho_{\mathrm{CDM}}$ & $0$ & $C(\varphi)q_{\mathrm{CDM}a}$ & $0$\\
$\varphi$ & $\frac{1}{2}\dot{\varphi}^{2} + V(\varphi)$ & $\frac{1}{2}\dot{\varphi}^{2} - V(\varphi)$ & $\dot{\varphi}\hat{\nabla}_{a}\varphi$ & $0$\\
\hline\hline
\end{tabular}
\end{center}
Throughout this paper an overdot denotes the derivative with
respect to the cosmic time $t$ and $\hat{\nabla}_{a}$ is the
covariant spatial derivative perpendicular to $u_{a}$ (up to first
order in perturbation). Note that it is the \emph{coupled} CDM
quantities which appear in the (background and perturbed) Einstein
equations, and also note that we have included the massless
neutrinos into the model here.

The five constraint equations in the model are given as
\begin{eqnarray}
0 &=&\hat{\nabla}^{c}(\epsilon_{\ \ cd}^{ab}u^{d}\varpi_{ab}); \\
\kappa q_{a} &=& -\frac{2\hat{\nabla}_{a}\theta}{3}+\hat{\nabla}^{b}\sigma_{ab}+\hat{\nabla}^{b}\varpi_{ab};\ \ \  \\
\mathcal{B}_{ab} &=&\left[
\hat{\nabla}^{c}\sigma_{d(a}+\hat{\nabla}^{c}\varpi_{d(a}\right] \epsilon_{b)ec}^{\ \ \ \ d}u^{e}; \\
\hat{\nabla}^{b}\mathcal{E}_{ab} &=&\frac{1}{2}\kappa \left[
\hat{\nabla}^{b}\pi_{ab}+ \frac{2}{3}\theta q_{a}+\frac{2}{3}\hat{\nabla}_{a}\rho\right]; \\
\hat{\nabla}^{b}\mathcal{B}_{ab} &=&\frac{1}{2}\kappa \left[
\hat{\nabla}_{c}q_{d}+(\rho+p)\varpi_{cd}\right]\epsilon_{ab}^{\ \
cd}u^{b}.
\end{eqnarray}
Here, $\epsilon_{abcd}$ is the covariant permutation tensor,
$\mathcal{E}_{ab}$ and $\mathcal{B}_{ab}$ are respectively the
electric and magnetic parts of the Weyl tensor
$\mathcal{W}_{abcd}$, given respectively through $\mathcal{E}_{ab}
= u^{c}u^{d}\mathcal{W}_{acbd}$ and $\mathcal{B}_{ab} = -
\frac{1}{2}u^{c}u^{d}\epsilon_{ac}^{\ \ ef}\mathcal{W}_{efbd}$.
$\theta, \sigma_{ab}, \varpi_{ab}$ come from the decomposition of
the covariant derivative of 4-velocity
\begin{eqnarray}
\nabla_{a}u_{b} &=& \sigma_{ab} + \varpi_{ab} + \frac{1}{3}\theta
h_{ab} + u_{a}A_{b}
\end{eqnarray}
with $A$ being the acceleration, $\theta = \nabla^{c}u_{c} =
3\dot{a}/a$ the expansion scalar, $\varpi_{ab} =
\hat{\nabla}_{[a}u_{b]}$ and $\sigma_{ab}$ the shear. Note that
$\theta$ in the above section has a completely different meaning.

In addition, the seven propagation equations are:
\begin{eqnarray}
\dot{\rho}+(\rho +p)\theta +\hat{\nabla}^{a}q_{a} &=&0; \\
\dot{q}_{a} + \frac{4}{3}\theta
q_{a}+(\rho +p)A_{a}-\hat{\nabla}_{a}p + \hat{\nabla}^{b}\pi _{ab} &=& 0; \nonumber\\
\label{eq:q_DM}\dot{q}_{a}+\frac{4}{3}\theta q_{a}+(\rho
+p)A_{a}-\hat{\nabla}_{a}p+\hat{\nabla}^{b}\pi _{ab}\nonumber\\ -
\frac{C_{\varphi}}{C}(\rho_{\mathrm{CDM}}\hat{\nabla}_{a}\varphi
-\dot{\varphi}q_{\mathrm{CDM}a}) &=& 0; \\
\label{eq:Raychaudhrui}\dot{\theta}+\frac{1}{3}\theta^{2} - \
\hat{\nabla}^{a}A_{a} + \frac{\kappa}{2}(\rho + 3p)  &=& 0; \\
\dot{\sigma}_{ab}+\frac{2}{3}\theta
\sigma_{ab}-\hat{\nabla}_{\langle a}A_{b\rangle}
+ \mathcal{E}_{ab}+\frac{1}{2}\kappa \pi_{ab} &=& 0; \\
\dot{\varpi}+\frac{2}{3}\theta\varpi - \hat{\nabla}_{[a}A_{b]} &=& 0; \\
\frac{1}{2}\kappa \left[\dot{\pi}_{ab} +
\frac{1}{3}\theta\pi_{ab}\right] - \frac{1}{2}\kappa \left[(\rho
+p)\sigma_{ab}\ +\hat{\nabla}_{\langle
a}q_{b\rangle }\right] \nonumber \\
-\left[ \dot{\mathcal{E}}_{ab} + \theta
\mathcal{E}_{ab}-\hat{\nabla}^{c}\mathcal{B}_{d(a}\epsilon_{b)ec}^{\
\ \ \
d}u^{e}\right]  &=& 0; \\
\dot{\mathcal{B}}_{ab} +
\theta\mathcal{B}_{ab}+\hat{\nabla}^{c}\mathcal{E}_{d(a}\epsilon_{b)ec}^{\
\ \ \ d}u^{e} + \frac{1}{2}\kappa
\hat{\nabla}^{c}\mathcal{\pi}_{d(a}\epsilon_{b)ec}^{\ \ \ \
d}u^{e} &=& 0.\ \ \ \
\end{eqnarray}%
where the angle bracket means taking the trace-free part of a
quantity. Note that the first of Eq.~(\ref{eq:q_DM}) is for normal
matter while the second is for the coupled dark matter.

Besides the above equations, it is useful to express the projected
Ricci scalar $\hat{R}$ into the hypersurfaces orthogonal to
$u^{a}$ as
\begin{eqnarray}\label{eq:projected_ricci_scalar}
\hat{R} &\doteq& 2\kappa\rho - \frac{2}{3}\theta ^{2}.
\end{eqnarray}%
The spatial derivative of the projected Ricci scalar,
$\eta_{a}\equiv \frac{1}{2}a\hat{\nabla}_{a}\hat{R}$, is given as
\begin{eqnarray}
\eta_{a} &=& a\kappa \hat{\nabla}_{a}\rho -
\frac{2a}{3}\theta\hat{\nabla}_{a}\theta,
\end{eqnarray}
and its propagation equation
\begin{eqnarray}
\dot{\eta}_{a} + \frac{2\theta }{3}\eta_{a} &=& -
\frac{2a}{3}\theta\hat{\nabla}_{a}\hat{\nabla}^{b}A_{b} - a\kappa
\hat{\nabla}_{a}\hat{\nabla}^{b}q_{b}.
\end{eqnarray}
As we are considering a spatially flat universe, the spatial
curvature must vanish on large scales which means that
$\hat{R}=0$. Thus, from Eq.~(\ref{eq:projected_ricci_scalar}), we
can obtain the Friedman equation
\begin{eqnarray}\label{eq:Friedman}
\frac{1}{3}\theta^{2} &=& \kappa\rho.
\end{eqnarray}
Note that in all the above equations $\rho,\ p,\ q_{a}$ and
$\pi_{ab}$ are all the total quantities contributed by all matter
species:
\begin{eqnarray}
\rho &=& \rho_{\gamma} + \rho_{\nu} + \rho_{\mathrm{B}} +
C(\varphi)\rho_{\mathrm{CDM}} + \frac{1}{2}\dot{\varphi}^{2} +
V(\varphi),\nonumber\\
p &=& \frac{1}{3}\rho_{\gamma} + \frac{1}{3}\rho_{\nu} +
p_{\mathrm{B}} + \frac{1}{2}\dot{\varphi}^{2} -
V(\varphi),\nonumber\\
q_{a} &=& q_{\gamma a} + q_{\nu a} + q_{\mathrm{B}a} +
C(\varphi)q_{\mathrm{CDM}a} +
\dot{\varphi}\hat{\nabla}_{a}\varphi,\nonumber\\
\pi_{ab} &=& \pi_{\gamma ab} + \pi_{\nu ab}.
\end{eqnarray}

Finally, there is the perturbed scalar field EOM
\begin{eqnarray}\label{eq:scalar_EOM}
\ddot{\varphi} + \theta\dot{\varphi} + \hat{\nabla}^{2}\varphi +
\frac{\partial V(\varphi)}{\partial\varphi} +
\rho_{\mathrm{CDM}}\frac{\partial C(\varphi)}{\partial\varphi} &=&
0.
\end{eqnarray}

\section{Equations in $k$ Space}

\label{appen:k-eqn}

To put the above perturbation equations into numerical
calculation, we need to write them in the $k$-space. As we want to
present the complete framework to study the structure formation in
coupled scalar field models, here we also list those equations.

First of all, the change into $k$-space is accomplished with the
aid of following Harmonic expansions:
\begin{eqnarray}
\mathcal{X}_{a} \equiv a\hat{\nabla}_{a}\rho =
\sum_{k}k\mathcal{X}Q^{k}_{a}\ \ \ \ \ q_{a} = \sum_{k}qQ^{k}_{a}\
\ \ \ \nonumber\\ \pi_{ab} = \sum_{k}\Pi Q^{k}_{ab}\ \ \ \ \
\mathcal{Z}_{a} \equiv a\hat{\nabla}_{a}\theta =
\sum_{k}\frac{k^{2}}{a}\mathcal{Z}Q^{k}_{a}\ \ \ \ \nonumber\\
\sigma_{ab} = \sum_{k}\frac{k}{a}\sigma Q^{k}_{ab}\ \ \ \ \
\eta_{a} =
\sum_{k}\frac{k^{3}}{a^{2}}\eta Q^{k}_{a}\nonumber\\
h_{a} \equiv \hat{\nabla}_{a}a = \sum_{k}khQ^{k}_{a}\ \ \ \ \
A_{a} = \sum_{k}\frac{k}{a}AQ^{k}_{a}\ \ \ \ \nonumber\\
\mathcal{E}_{ab} = - \sum_{k}\frac{k^{2}}{a^{2}}\phi Q^{k}_{ab}\ \
\ \ \ \hat{\nabla}_{a}\varphi = \sum_{k}\frac{k}{a}\xi Q^{k}_{a}
\end{eqnarray}
where $Q^{k}_{a} = \frac{a}{k}\hat{\nabla}_{a}Q^{k}$ with $Q^{k}$
being the zero order eigenfunctions of the comoving Laplacian
$a^{2}\hat{\nabla}^{2}$ ($a^{2}\hat{\nabla}^{2}Q^{k} =
k^{2}Q^{k}$) and $Q^{k}_{ab} = \frac{a}{k}\hat{\nabla}_{\langle
a}Q^{k}_{b\rangle}$. Note that we only consider scalar mode
perturbations in the present paper and so shall neglect the second
order quantities such as $\mathcal{B}_{ab}$ and $\varpi_{ab}$.

With these we list the equations that will be used in the
numerical calculation of the linear large scale structure.

\subsubsection{Background Evolution for Energy Densities}

\begin{eqnarray}
\rho_{\gamma}' + 4\frac{a'}{a}\rho_{\gamma} &=& 0,\\
\rho_{\nu}' + 4\frac{a'}{a}\rho_{\nu} &=& 0,\\
\rho_{\mathrm{B}}' + 3\frac{a'}{a}\rho_{\mathrm{B}} &=& 0,\\
\label{eq:bg_rho_CDM}\rho_{\mathrm{CDM}}' +
3\frac{a'}{a}\rho_{\mathrm{CDM}} &=& 0.
\end{eqnarray}
Note that because our definition of the dark matter energy density
includes only the mass density, but not the contribution from its
coupling to the scalar field, so $\rho_{\mathrm{CDM}}$ evolves as
in $\Lambda$CDM. One can also understand this as follows: the
coupling to the scalar field produces a fifth force on the dark
matter particles, making their trajectories be non-geodesic, but
as we shall see below, the fifth force is spatial (perpendicular
to the worldline of the dark matter particle) and cannot change
the mass of dark matter particles. Consequently the averaged dark
matter mass density is the same as in $\Lambda$CDM. One can of
course define the energy momentum tensor for dark matter as
including $C(\varphi)$, and in this case Eq.~(\ref{eq:bg_rho_CDM})
will no longer be valid and we end up with varying mass dark
matter particles.

\subsubsection{Propagations of Spatial Gradients of Densities}

\begin{eqnarray}
\Delta'_{\gamma} + \frac{4}{3}k\mathcal{Z} - 4\frac{a'}{a}A +
kv_{\gamma} &=& 0,\\
\Delta'_{\nu} + \frac{4}{3}k\mathcal{Z} - 4\frac{a'}{a}A +
kv_{\nu} &=& 0,\\
\Delta'_{\mathrm{B}} + \left(k\mathcal{Z} - 3\frac{a'}{a}A  +
kv_{_{\mathrm{B}}}\right)
+ 3\frac{a'}{a}c^{2}_{s}\Delta_{\mathrm{B}} &=& 0,\\
\Delta'_{\mathrm{CDM}} + k\mathcal{Z} - 3\frac{a'}{a}A +
kv_{\mathrm{CDM}} &=& 0.
\end{eqnarray}

\subsubsection{Propagations of Heat Fluxes}

\begin{eqnarray}
v'_{\nu} + \frac{k}{3}\left(2\frac{\Pi_{\nu}}{\rho_{\nu}} + 4A -
\Delta_{\nu}\right) &=& 0,\\
v'_{\gamma} + \frac{k}{3}\left(2\frac{\Pi_{\gamma}}{\rho_{\gamma}}
+ 4A - \Delta_{\gamma}\right)\nonumber\\ +
an_{e}\sigma_{\mathrm{T}}
\left(v_{\gamma} - \frac{4}{3}v_{\mathrm{B}}\right) &=& 0,\\
v'_{\mathrm{B}} + \frac{a'}{a}(1-3c^{2}_{s})v_{\mathrm{B}} +
kA\nonumber\\ - kc^{2}_{s}\Delta_{\mathrm{B}} +
an_{e}\sigma_{\mathrm{T}}\frac{\rho_{\gamma}}{\rho_{\mathrm{B}}}
\left(\frac{4}{3}v_{\mathrm{B}} - v_{\gamma}\right) &=& 0,\\
\label{eq:pb_v_CDM}v'_{\mathrm{CDM}} +
\frac{a'}{a}v_{\mathrm{CDM}} + kA +
\frac{C_{\varphi}}{C}\varphi'v_{\mathrm{CDM}} -
\frac{C_{\varphi}}{C}k\xi &=& 0.
\end{eqnarray}

\subsubsection{Other Propagation Equations}

\begin{eqnarray}
k\left(\sigma' + \frac{a'}{a}\sigma\right) - k^{2}\left(A +
\phi\right) +
\frac{1}{2}\kappa\Pi a^{2} &=& 0,\\
k^{2}\left(\phi' + \frac{a'}{a}\phi\right)\nonumber\\ +
\frac{1}{2}\kappa \left[\Pi' + \frac{a'}{a}\Pi - k(\rho + p)\sigma
- kq\right]a^{2}
&=& 0,\\
k\eta' + 2\frac{a'}{a}kA + \kappa qa^{2} &=& 0,
\end{eqnarray}
in which we have used
\begin{eqnarray}
\rho &=& \rho_{\gamma} + \rho_{\nu} + \rho_{\mathrm{B}} +
C(\varphi)\rho_{\mathrm{CDM}} + \frac{1}{2a^{2}}\varphi'^{2} +
V(\varphi),\ \ \ \ \\
p &=& \frac{1}{3}\rho_{\gamma} +
\frac{1}{3}\rho_{\nu} +
\frac{1}{2a^{2}}\varphi'^{2} - V(\varphi),\\
q &=& \rho_{\gamma}v_{\gamma} + \rho_{\nu}v_{\nu} +
\rho_{\mathrm{B}}v_{\mathrm{B}} +
\frac{k}{a^{2}}\varphi'\xi,\ \ \\
\Pi &=& \Pi_{\gamma} + \Pi_{\nu},\\
\mathcal{X} &=& \rho_{\gamma}\Delta_{\gamma} +
\rho_{\nu}\Delta_{\nu} +
\rho_{_{\mathrm{B}}}\Delta_{_{\mathrm{B}}} +
C(\varphi)\rho_{_{\mathrm{CDM}}}\Delta_{_{\mathrm{CDM}}}\nonumber\\
&& + \frac{1}{a^{2}}\varphi'\xi' +
\left(\frac{1}{a^{2}}\frac{C_{\varphi}}{C}\varphi'^{2} +
V_{\varphi} + C_{\varphi}\rho_{\mathrm{CDM}}\right)\xi\ \
\end{eqnarray}
and defined the density contrast
$\Delta_{i}=\mathcal{X}_{i}/\rho_{i}$ for matter species $i$.

\subsubsection{Constraint Equations}

\begin{eqnarray}
k^{2}(\mathcal{Z} - \sigma) + \frac{3}{2}\kappa qa^{2} &=& 0,\\
k^{3}\phi + \frac{1}{2}\kappa\left[k(\Pi + \mathcal{X}) +
3\frac{a'}{a}q\right]a^{2} &=& 0,\\
k^{2}\eta + 2\frac{a'}{a}k\mathcal{Z} - \kappa\mathcal{X}a^{2} &=&
0.
\end{eqnarray}

\subsubsection{Scalar Field Equations of Motion}

\begin{eqnarray}
\label{eq:sfbgd_EOM}\varphi'' + 2\frac{a'}{a}\varphi' +
\frac{\partial V(\varphi)}{\partial\varphi}a^{2} +
\rho_{\mathrm{CDM}}\frac{\partial
C(\varphi)}{\partial\varphi}a^{2} &=& 0,\\
\label{eq:sfpert_EOM}\xi'' + 2\frac{a'}{a}\xi' + \left(k^{2} +
a^{2}V_{\varphi\varphi}
+ a^{2}\rho_{\mathrm{CDM}}C_{\varphi\varphi}\right)\xi\nonumber\\
+ \left(2\varphi'' + \frac{a'}{a}\varphi'\right)A
+ a^{2}C_{\varphi}\rho_{\mathrm{CDM}}\Delta_{\mathrm{CDM}}\nonumber\\
+ \left(k\mathcal{Z} + A'\right)\varphi' &=& 0.
\end{eqnarray}

\subsubsection{The Friedmann Equations}

\begin{eqnarray}
3\left(\frac{a'}{a}\right)^{2} &=& \kappa\rho a^{2},\\
\frac{a''}{a} - \left(\frac{a'}{a}\right)^{2} &=& -
\frac{\kappa}{6}(\rho + 3p)a^{2}.
\end{eqnarray}

When it comes to perturbation calculations, we need to fix a
gauge, \emph{i.e.}, choose a $u_{a}$. One possibility is to use
the 4-velocity of dark matter particles as our $u_{a}$. In this
case the dark matter heat flux is zero and according to
Eq.~(\ref{eq:pb_v_CDM}) we will simply have
$A=\frac{C_{\varphi}}{C}\xi$, and this relation can then be used
to replace the $A$'s appearing in all the above equations. Another
possibility is to choose $u_{a}$ such that $A=0$: in this case
$v_{\mathrm{CDM}}$ will become nonzero and we need to dynamically
evolve it. Below in the numerical calculations we shall adopt the
second possibility.

\section{Discretized Equations for N-body Simulations Non-linear Regime}

\label{appen:discret}

In the MLAPM code the partial differential equation
Eq.~(\ref{eq:INTPoisson}) is (and in our modified code
Eq.~(\ref{eq:INTphiEOM}) will also be) solved on discretized grid
points, and as such we must develop the discretized versions of
Eqs.~(\ref{eq:INTdxdtcomov} - \ref{eq:INTphiEOM}) to be
implemented into the code.

But before going on to the discretization, we need to address a
technical issue. As the potential is highly nonlinear, in the high
density regime the value of the scalar field
$\sqrt{\kappa}\varphi$ will be very close to 0, and this is
potentially a disaster as during the numerical solution process
the value of $\sqrt{\kappa}\varphi$ might easily go into the
forbidden region $\varphi<0$ \cite{Oyaizu2008}. One way of solving
this problem is to define $\chi=\bar{\chi}e^{u}$ in which
$\bar{\chi}$ is the background value of $\chi$, as in
\cite{Oyaizu2008}. Then the new variable $u$ takes value in
$(-\infty, \infty)$ so that $e^{u}$ is positive definite which
ensures that $\chi>0$. However, since there are already
exponentials of $\chi$ in the potential, this substitution will
result terms involving $\exp\left[\exp(u)\right]$, which could
potentially magnify any numerical error in $u$.

Instead, we can define a new variable $u$ according to
\begin{eqnarray}
e^{u}+1 &=& e^{\chi}.
\end{eqnarray}
By this, $u$ still takes value in $(-\infty, \infty)$,
$e^{u}\in(0, \infty)$ and thus $e^{\chi}\in(1, \infty)$ which
ensures that $\chi$ is positive definite in numerical solutions.
Besides, $e^{\beta\chi}=\left[1+e^{u}\right]^{\beta}$ so that
there will be no exponential-of-exponential terms, and the only
exponential is what we have for the potential itself. $\beta=-1$
above.

Then the Poisson equation becomes
\begin{eqnarray}\label{eq:u_Poisson}
\nabla^{2}\Phi_{c} &=&
\frac{3}{2}\Omega_{\mathrm{CDM}}\left[\rho_{c}
\left(1+e^{u}\right)^{\gamma} -
e^{\gamma\sqrt{\kappa}\bar{\varphi}}\right]\nonumber\\
&& -
\frac{3\Omega_{V_{0}}a^{3}}{\left[1-\left(1+e^{u}\right)^{\beta}\right]^{\mu}}
+ 3\bar{\Omega}_{V}a^{3},
\end{eqnarray}
where we have defined $\bar{\Omega}_{V}=\kappa
V(\bar{\varphi})/3H_{0}^{2}$ which is determined by background
cosmology, the quantity $e^{\gamma\sqrt{\kappa}\bar{\varphi}}$ is
also determined solely by background cosmology. These background
quantities should not bother us here.

The scalar field EOM becomes
\begin{eqnarray}\label{eq:u_phi_EOM}
&&\frac{ac^{2}}{\left(H_{0}B\right)^{2}}\nabla\cdot\left(\frac{e^{u}}{1+e^{u}}\nabla
u\right)\nonumber\\
&=& 3\gamma\Omega_{\mathrm{CDM}}\rho_{c}
\left(1+e^{u}\right)^{\gamma} +
\frac{3\mu\beta\Omega_{V_{0}}a^{3}\left(1+e^{u}\right)^{\beta}}
{\left[1-\left(1+e^{u}\right)^{\beta}\right]^{\mu+1}}\nonumber\\
&& -
3\gamma\Omega_{\mathrm{CDM}}e^{\gamma\sqrt{\kappa}\bar{\varphi}} -
\frac{3\mu\beta\Omega_{V_{0}}a^{3}e^{\beta\sqrt{\kappa}\bar{\varphi}}}
{\left[1-e^{\beta\sqrt{\kappa}\bar{\varphi}}\right]^{m+1}}
\end{eqnarray}
in which we have used the fact that
$\chi=\log(1+e^{u})\Rightarrow\nabla\chi=\frac{e^{u}}{1+e^{u}}\nabla
u$, and moved all terms depending only on background cosmology
(the source terms) to the right hand side.

So, in terms of the new variable $u$, the set of equations used in
the N-body code should be
\begin{eqnarray}\label{eq:u_dxdt}
\frac{d\mathbf{x}_{c}}{dt_{c}} &=& \frac{\mathbf{p}_{c}}{a^{2}},\\
\label{eq:u_dpdt} \frac{d\mathbf{p}_{c}}{dt_{c}} &=&
-\frac{1}{a}\nabla\Phi_{c} -
\frac{c^{2}\gamma}{\left(H_{0}B\right)^{2}}\frac{e^{u}}{1+e^{u}}\nabla
u
\end{eqnarray}
plus Eqs.~(\ref{eq:u_Poisson}, \ref{eq:u_phi_EOM}). These
equations will ultimately be used in the code. Among them,
Eqs.~(\ref{eq:u_Poisson}, \ref{eq:u_dpdt}) will use the value of
$u$ while Eq.~(\ref{eq:u_phi_EOM}) solves for $u$. In order that
these equations can be integrated into MLAPM, we need to
discretize Eq.~(\ref{eq:u_phi_EOM}) for the application of
Newton-Gauss-Seidel iterations.

To discretize Eq.~(\ref{eq:u_phi_EOM}), let us define
$b\equiv\frac{e^{u}}{1+e^{u}}$. The discretization involves
writing down a discretion version of this equation on a uniform
grid with grid spacing $h$. Suppose we require second order
precision as is in the standard Poisson solver of MLAPM, then
$\nabla u$ in one dimension can be written as
\begin{eqnarray}
\nabla u &\rightarrow& \nabla^{h}u_{j}\ =\
\frac{u_{j+1}-u_{j-1}}{2h}
\end{eqnarray}
where a subscript $_{j}$ means that the quantity is evaluated on
the $j$-th point. Of course the generalization to three dimensions
is straightforward.

The factor $b$ in $\nabla\cdot\left(b\nabla u\right)$ makes this a
standard variable coefficient problem. We need also discretize
$b$, and do it in this way (again for one dimension):
\begin{widetext}
\begin{eqnarray}
\nabla\cdot\left(b\nabla u\right) &\rightarrow&
\left(\nabla^{h}b_{j}\right)\cdot\left(\nabla^{h}u_{j}\right) +
b_{j}\nabla^{h2}u_{j}\nonumber\\
&=& \frac{b_{j+1/2}-b_{j-1/2}}{h}\frac{u_{j+1}-u_{j-1}}{2h} +
\frac{b_{j+1/2}+b_{j-1/2}}{2}\frac{u_{j+1}-2u_{j}+u_{j-1}}{h^{2}}\nonumber\\
&=& \frac{1}{h^{2}}\left[b_{j+\frac{1}{2}}u_{j+1} -
u_{j}\left(b_{j+\frac{1}{2}}+b_{j-\frac{1}{2}}\right) +
b_{j-\frac{1}{2}}u_{j-1}\right]
\end{eqnarray}
\end{widetext}
where we have defined
$b_{j+\frac{1}{2}}=\left(b_{j}+b_{j+1}\right)/2$ and
$b_{j-\frac{1}{2}}=\left(b_{j-1}+b_{j}\right)/2$. This can be
easily generalize to three dimensions as
\begin{widetext}
\begin{eqnarray}
\nabla\cdot\left(b\nabla u\right) &\rightarrow&
\frac{1}{h^{2}}\left[b_{i+\frac{1}{2},j,k}u_{i+1,j,k} -
u_{i,j,k}\left(b_{i+\frac{1}{2},j,k}+b_{i-\frac{1}{2},j,k}\right)
+ b_{i-\frac{1}{2},j,k}u_{i-1,j,k}\right]\nonumber\\
&& + \frac{1}{h^{2}}\left[b_{i,j+\frac{1}{2},k}u_{i,j+1,k} -
u_{i,j,k}\left(b_{i,j+\frac{1}{2},k}+b_{i,j-\frac{1}{2},k}\right)
+ b_{i,j-\frac{1}{2},k}u_{i,j-1,k}\right]\nonumber\\
&& + \frac{1}{h^{2}}\left[b_{i,j,k+\frac{1}{2}}u_{i,j,k+1} -
u_{i,j,k}\left(b_{i,j,k+\frac{1}{2}}+b_{i,j,k-\frac{1}{2}}\right)
+ b_{i,j,k-\frac{1}{2}}u_{i,j,k-1}\right].
\end{eqnarray}
\end{widetext}
Then the discrete version of Eq.~(\ref{eq:u_phi_EOM}) is
\begin{eqnarray}\label{eq:diffop}
L^{h}\left(u_{i,j,k}\right) &=& 0,
\end{eqnarray}
in which
\begin{widetext}
\begin{eqnarray}
L^{h}\left(u_{i,j,k}\right) &=&
\frac{1}{h^{2}}\left[b_{i+\frac{1}{2},j,k}u_{i+1,j,k} -
u_{i,j,k}\left(b_{i+\frac{1}{2},j,k}+b_{i-\frac{1}{2},j,k}\right)
+ b_{i-\frac{1}{2},j,k}u_{i-1,j,k}\right]\nonumber\\
&& + \frac{1}{h^{2}}\left[b_{i,j+\frac{1}{2},k}u_{i,j+1,k} -
u_{i,j,k}\left(b_{i,j+\frac{1}{2},k}+b_{i,j-\frac{1}{2},k}\right)
+ b_{i,j-\frac{1}{2},k}u_{i,j-1,k}\right]\nonumber\\
&& + \frac{1}{h^{2}}\left[b_{i,j,k+\frac{1}{2}}u_{i,j,k+1} -
u_{i,j,k}\left(b_{i,j,k+\frac{1}{2}}+b_{i,j,k-\frac{1}{2}}\right)
+ b_{i,j,k-\frac{1}{2}}u_{i,j,k-1}\right]\nonumber\\
&&-\frac{\left(H_{0}B\right)^{2}}{ac^{2}}\left[3\gamma\Omega_{\mathrm{CDM}}\rho_{c,i,j,k}
\left(1+e^{u_{i,j,k}}\right)^{\gamma} +
\frac{3\mu\beta\Omega_{V_{0}}a^{3}\left(1+e^{u_{i,j,k}}\right)^{\beta}}
{\left[1-\left(1+e^{u_{i,j,k}}\right)^{\beta}\right]^{\mu+1}}\right]\nonumber\\
&& +\frac{\left(H_{0}B\right)^{2}}{ac^{2}}\left[
3\gamma\Omega_{\mathrm{CDM}}e^{\gamma\sqrt{\kappa}\bar{\varphi}} +
\frac{3\mu\beta\Omega_{V_{0}}a^{3}e^{\beta\sqrt{\kappa}\bar{\varphi}}}
{\left[1-e^{\beta\sqrt{\kappa}\bar{\varphi}}\right]^{\mu+1}}\right].
\end{eqnarray}
\end{widetext}
Then the Newton-Gauss-Seidel iteration says that we can obtain a
new (and often more accurate) solution of $u$,
$u^{\mathrm{new}}_{i,j,k}$, using our knowledge about the old (and
less accurate) solution $u^{\mathrm{old}}_{i,j,k}$ as
\begin{eqnarray}\label{eq:GS}
u^{\mathrm{new}}_{i,j,k} &=& u^{\mathrm{old}}_{i,j,k} -
\frac{L^{h}\left(u^{\mathrm{old}}_{i,j,k}\right)}{\partial
L^{h}\left(u^{\mathrm{old}}_{i,j,k}\right)/\partial u_{i,j,k}}.
\end{eqnarray}
The old solution will be replaced by the new solution to
$u_{i,j,k}$ once the new solution is ready, using the red-black
Gauss-Seidel sweeping scheme. Note that
\begin{widetext}
\begin{eqnarray}
\frac{\partial L^{h}(u_{i,j,k})}{\partial u_{i,j,k}} &=&
\frac{1}{2h^{2}}\frac{e^{u_{i,j,k}}}{\left(1
+e^{u_{i,j,k}}\right)^{2}}\left[u_{i+1,j,k}+u_{i-1,j,k}+u_{i,j+1,k}
+u_{i,j-1,k}+u_{i,j,k+1}+u_{i,j,k-1}-6u_{i,j,k}\right]\nonumber\\
&&-\frac{1}{2h^{2}}\left[b_{i+1,j,k}+b_{i-1,j,k}+b_{i,j+1,k}
+b_{i,j-1,k}+b_{i,j,k+1}+b_{i,j,k-1}+6b_{i,j,k}\right]\nonumber\\
&&-\frac{\left(H_{0}B\right)^{2}}{ac^{2}}3\gamma^{2}\Omega_{\mathrm{CDM}}\rho_{c,i,j,k}
\left(1+e^{u_{i,j,k}}\right)^{\gamma}b_{i,j,k}\nonumber\\
&&-\frac{\left(H_{0}B\right)^{2}}{ac^{2}}
\frac{3\mu\beta^{2}\Omega_{V_{0}}a^{3}\left(1+e^{u_{i,j,k}}\right)^{\beta}}
{\left[1-\left(1+e^{u_{i,j,k}}\right)^{\beta}\right]^{\mu+1}}b_{i,j,k}
\left[1+(\mu+1)\frac{\left(1+e^{u_{i,j,k}}\right)^{\beta}}{1-\left(1+e^{u_{i,j,k}}\right)^{\beta}}\right].
\end{eqnarray}
\end{widetext}

In principle, if we start from a high redshift, then the initial
guess of $u_{i,j,k}$ could be such that the initial value of
$\chi$ in all the space is equal to the background value
$\bar{\chi}$, because anyway at this time we expect this to be
approximately true. For subsequent time steps we could use the
solution for $u_{i,j,k}$ at the previous time step as our initial
guess; if the time step is small enough then we do not expect $u$
to change significantly between consecutive times so that such a
guess will be good enough for the iteration to converge fast.

In practice, however, due to specific features and algorithm of
the MLAPM code \cite{MLAPM}, the above procedure may be slightly
different in details.

\end{document}